\documentclass[%
  reprint,
  superscriptaddress,
  preprintnumbers,
  nobibnotes,
  amsmath,amssymb,
  prb,
  floatfix,
  ]{revtex4-2}

  \usepackage{graphicx}% Include figure files
  \usepackage{dcolumn}% Align table columns on decimal point
  \usepackage{tabularx}
  \usepackage{booktabs}

  \usepackage{bm}
  \usepackage{xstring}
  \usepackage{xcolor}
  \usepackage{hyperref}
  \usepackage{multirow}

  \usepackage{mathtools}

\begin{document}

  \title{ Born effective charges and vibrational spectra in super and bad conducting metals }
  \author{Guglielmo Marchese}
  \affiliation{Dipartimento di Fisica, Università di Roma La Sapienza, Piazzale Aldo Moro 5, I-00185 Roma, Italy}

  \author{Francesco Macheda}
  \affiliation{Istituto Italiano di Tecnologia, Graphene Labs, Via Morego 30, I-16163 Genova, Italy}

  \author{Luca Binci}
  \affiliation{Theory and Simulation of Materials (THEOS), and National Centre for Computational Design and Discovery of Novel Materials (MARVEL),
  École Polytechnique Fédérale de Lausanne (EPFL), CH-1015 Lausanne, Switzerland}

  \author{Matteo Calandra}
  \affiliation{Department of Physics, University of Trento, Via Sommarive 14, 38123 Povo, Italy}
   \affiliation{Istituto Italiano di Tecnologia, Graphene Labs, Via Morego 30, I-16163 Genova, Italy}

  \author{Paolo Barone}
  \affiliation{  Dipartimento di Fisica, Università di Roma La Sapienza, Piazzale Aldo Moro 5, I-00185 Roma, Italy}
  \affiliation{ Consiglio Nazionale delle Ricerche, Institute for Superconducting and other Innovative Materials and Devices (CNR-SPIN), Area della Ricerca di Tor Vergata, Via del Fosso del Cavaliere 100,
  I-00133 Rome, Italy }

  \author{Francesco Mauri}
  \email{francesco.mauri@uniroma1.it}
  \affiliation{Dipartimento di Fisica, Università di Roma La Sapienza, Piazzale Aldo Moro 5, I-00185 Roma, Italy}
  \affiliation{Istituto Italiano di Tecnologia, Graphene Labs, Via Morego 30, I-16163 Genova, Italy}

  \date{\today}% It is always \today, today,

\begin{abstract}
Interactions mediated by electron-phonon coupling are responsible for important
cooperative phenomena in metals such as superconductivity and charge-
density waves. The same interaction mechanisms produce strong collision rates
in the normal phase of correlated metals, causing sizeable reductions of the dc
conductivity and reflectivity. As a consequence, low-energy excitations like
phonons, which are crucial for materials characterization, become visible in
optical infrared spectra. A quantitative assessment of vibrational resonances
requires the evaluation of dynamical Born effective charges, which quantify the
coupling between macroscopic electric fields and lattice deformations. We show
that the Born effective charges of metals crucially depend on the collision
regime of conducting electrons. In particular, we describe, within a first
principles framework, the impact of electron scattering on the infrared
vibrational resonances, from the undamped, collisionless  regime to the
overdamped, collision-dominated limit. Our approach enables the interpretation
of vibrational reflectance measurements of both super and bad conducting
metals, as we illustrate for the case of strongly electron-phonon coupled
superhydride H$_3$S.
\end{abstract}

  \maketitle
  
  \renewcommand{\figurename}{}
  \renewcommand{\thefigure}{\textbf{Fig.~\arabic{figure}}}

\noindent \textbf{\large Introduction}\\  
Optical spectroscopy is a widespread and informative technique for characterizing condensed-matter systems through the analysis of their elementary excitations. In conventional metals, i.e., crystalline systems of independent electrons, the optical response is mostly contributed by free charge carriers, that perfectly reflect the incoming electromagnetic radiation for frequencies up to the plasma frequency, typically falling in the ultraviolet range. Low-energy excitations, such as phonons and spin fluctuations, are therefore hardly discernible in the infrared (IR) reflectivity spectra of most conventional metals \cite{bruesch_phonons_1986}, while they become accessible in weakly doped semiconductors with low densities of charge carriers \cite{yu2010fundamentals}, or semimetals like graphene and graphite \cite{kuzmnenko_prl2008,PhysRevLett.103.116804,Cappelluti_bigraphene_prb2012, binci_first-principles_2021}. The free-carrier contribution to the optical response can also be heavily modified by the presence of strong scattering mechanisms and many-body interactions, whose effects can be described in a quasiparticle picture by the self-energy of the interacting Green's function, encompassing both a mass enhancement and, more importantly, a reduced lifetime (strong damping) of electronic particle-like excitations. This is the case for several correlated systems such as high-T$_c$ cuprates, pnictides, transition-metal compounds and heavy-fermion systems, generally displaying anomalous metallic properties in the normal phase preceding the onset of cooperative long-range ordered electronic states. Indeed, the analysis of their electromagnetic response from the terahertz to the optical range has revealed a rich spectrum of low-energy excitations reflecting the complex nature of the underlying strongly correlated phenomena \cite{Basov_RMP2005,Basov_pnictides_natphys2009,Basov_RMP2011}. Particularly interesting are those compounds where a strong electron-phonon (e-ph) interaction leads to the onset of charge-density wave (CDW) and/or superconducting (SC) state, such as transition-metal dichalcogenides \cite{Wilson_TMDrev1975, KLEMM_TMDrev2015, CastroNeto_prl2001} or superhydrides, hydrogen-rich compounds displaying record SC critical temperatures under ultra-high pressures \cite{drozdov_conventional_2015, Drozdov2019, Somayazulu2019, FLORESLIVAS20201}. 
The strong electron-phonon scattering results in a significant reduction of the free-carrier contribution to the dielectric response, such that the low-energy spectral features, including phonons, become accessible to optical spectroscopy. If this scenario is realized, IR reflectivity measurements would prove a powerful and non-invasive tool especially for the characterization of high-pressure hydrides and metallic hydrogen, providing direct access to their vibrational properties as well as confirming the conventional pairing mechanism for superconductivity mediated by strong e-ph coupling \cite{capitani_spectroscopic_2017, Carbotte_prl2018,carbotte_spectroscopic_2019,Loubeyre_nature2020}. 

This opportunity calls for a reexamination of the vibrational contribution to the dielectric response in different scattering regimes. Its description requires the evaluation of the oscillator strengths associated with each optically active phonon \cite{Born1988dynamical,de_gironcoli_piezoelectric_1989,Gonze_prb1997}. In crystalline insulators, these are expressed in terms of Born effective charges (BECs), measuring the change of electric polarization induced by an unitary atomic displacement under the condition of zero macroscopic electric field \cite{de_gironcoli_piezoelectric_1989,Ghosez_prb1998}. Such condition is then translated in the use of BECs for the calculation of the transverse macroscopic dielectric response. A first-principles generalization to metals has been proposed only recently by introducing frequency-dependent, nonadiabatic BECs \cite{bistoni_giant_2019,binci_first-principles_2021,dreyer_nonadiabatic_2022,PhysRevB.106.L180303}, that implement the so-called charged-phonon effect discussed in the past within the context of fullerenes \cite{Rice_prb1992} and more recently of graphene-based systems \cite{PhysRevLett.103.116804, Cappelluti_bigraphene_prb2012}. However, such nonadiabatic BECs, that we name in the present work as `dynamical' BECs, have been derived under the stringent hypothesis that the electrons are undamped quasiparticles with an infinite lifetime.  Interestingly, Ref. \cite{dreyer_nonadiabatic_2022} identifies a dynamical sumrule relating the dynamical BECs to the Drude weight and the optical conductivity. As in metals the latter is known to be critically affected by the electron-lifetime, the impact of electron damping on BECs remains to be studied. We address precisely this problem within a linear-response \textit{ab-initio} approach rooted in Density Functional Perturbation Theory (DFPT) \cite{Gonze_prb1997, baroni_phonons_2001} that allows us to model, under controlled approximations and on an equal footing, dynamical and finite-lifetime effects on both the electronic and vibrational contributions to the dielectric response of metals in different scattering regimes. As a paradigmatic example, we apply our approach to simulate the IR reflectivity spectrum of H$_3$S at 150 GPa measured by Capitani et al. \cite{capitani_spectroscopic_2017}, corroborating the assignment of a spectroscopical signature at $\sim 150~$ meV to a vibrational resonance of the cubic $Im\bar{3}m$ phase. \\

\noindent \textbf{\large Infrared reflectivity and macroscopic dielectric response}\\  
The reflectivity spectrum $R(\omega,\Sigma)$ of an isotropic crystal illuminated by an electromagnetic radiation of frequency $\omega$ can be related to its complex dielectric function $\varepsilon(\omega,\Sigma)$ and to the refractive index of the environment $n_0$ via the Fresnel equation, that for normal incidence reads:
  \begin{align}
    R(\omega,\Sigma) = \left\lvert \frac{ \sqrt{\varepsilon(\omega,\Sigma)} - n_0}{ \sqrt{\varepsilon(\omega,\Sigma)} + n_0} \right\rvert^2,
    \label{eq_main:R}
  \end{align}
  where we emphasize the dependence on the single-particle self-energies $\Sigma$ of both electrons (as referred to the Kohn-Sham system) and phonons. The dielectric response can be decomposed into electronic and lattice contributions as: 
  \begin{align}
    \varepsilon(\omega,\Sigma) = 1 + 4\pi\chi^{\text{el}}(\omega,\Sigma^{\text{el}}) +  4\pi\chi^{\text{vibr}}(\omega,\Sigma^{\text{vibr}},\Sigma^\text{el}).
    \label{eq:epsilontot}
  \end{align}   
  The electronic contribution, originating from intraband (Drude-like) and interband transitions, may be further decomposed as $\chi^{\text{el}}(\omega,\Sigma^{\text{el}})=\frac{1}{-i\omega}\sigma_{\text{Drude}}(\omega,\Sigma^{\text{el}}) +  \frac{1}{-i\omega}\sigma_{\text{Inter}}(\omega,\Sigma^{\text{el}})$, where $\sigma(\omega)$ indicates the optical conductivity. In \ref{fig:R_excitations}\textbf{a}, we show the total reflectivity spectrum of the normal phase of cubic H$_3$S, with the lattice contribution evaluated as discussed below and including the electronic contributions at different levels of theory. The red curve, labeled as `Undamped Drude’, is obtained with the assumption of vanishing electron scattering rates (i.e., $\Sigma^{\text{el}}=0$), and neglecting interband contributions: as anticipated, the Drude-like contribution of free charges perfectly reflects the electromagnetic radiation up to a plasma frequency $\omega_{\text{p}}\sim 13.3~$eV, precluding access to any lower-energy elementary excitation such as phonons. We relax  the independent-electron approximation via the extended-Drude model \cite{nam_theory_1967}(Methods, Supplementary Section \ref{si:extdrude}). The resulting finite electron linewidths and lineshifts induced by inelastic electron-phonon and  elastic electron-impurity scattering (i.e., $\Sigma^{\text{el}} \neq 0$) produce a significant reduction of reflectivity in a broad frequency range, displayed as a grey curve (‘Damped Drude’). Such reduction provides access to low energy excitations even in the far-IR window, as exemplified by the two sharp vibrational resonances highlighted by vertical arrows. Inclusion of interband contributions (black curve) further reduces the reflectivity but at frequencies higher than the vibrational features, suggesting that a description restricted to electronic states lying near the Fermi surface (a Landau Fermi liquid theory) is sufficient.
  Having assessed the experimental access to IR excitations in metals beyond the undamped regime, we focus on the vibrational contribution to the dielectric response, that can be modelled as \cite{bistoni_giant_2019, binci_first-principles_2021} (Supplementary Section \ref{si:vibr}):
  \begin{align}
    \chi^{\text{vibr}}_{\alpha \beta}(\omega,\Sigma^{\text{vibr}},\Sigma^{\text{el}}) &= 
    \frac{e^2}{\Omega}\sum_{\mu}^{\rm modes} \frac{d^{\mu}_{\alpha}(\omega,\Sigma^{\text{el}})d^{\mu}_{\beta}(\omega,\Sigma^{\text{el}})}{\omega_\mu^2 - \left( \omega + i\gamma_\mu/2 \right)^2}.  
    \label{eq:chivibr}
  \end{align} 
  Here $\alpha, \beta$ are Cartesian indices, $\Omega$ is the unit-cell volume, $e$ is the electron charge, $\omega_{\mu}$ and $\hbar \gamma_\mu=-2\mathrm{Im}(\Sigma^{\text{vibr}}_{\mu})$ are the frequency and the linewidth 
  %in the static limit 
  of a phonon mode $\mu$,  respectively, and $d^{\mu}_{\alpha}(\omega,\Sigma^{\text{el}})$ is the associated oscillator strength. 
  %{\color{blue} (see Suppl. Table \ref{tab:h3s_al_vel_vs_p})}. 
  The latter is defined as \cite{binci_first-principles_2021, bistoni_giant_2019} $d^{\mu}_{\alpha}(\omega,\Sigma^{\text{el}}) =  \sum_{s\beta} \bar{Z}_{s,\alpha\beta}(\omega,\Sigma^{\text{el}})    e^{\mu}_{s\beta}/ \sqrt{M_{s}}$, where $e^{\mu}_{s\beta}$ is the vector of atomic displacements for the phonon of mode $\mu$, $M_{s}$ denotes the ionic mass of an atom $s$ within the unit cell and $\bar{Z}_{s,\alpha\beta}(\omega,\Sigma^{\text{el}})$ are the BEC tensors, that are in general frequency-dependent complex quantities \cite{bistoni_giant_2019,binci_first-principles_2021,dreyer_nonadiabatic_2022,PhysRevB.106.L180303}; here, we again  {\color{black} stress} their dependence on the electronic quasiparticle self-energy. As they are key quantities for assessing the strength of the vibrational response, in the following we analyse in more details dynamical BECs of metallic systems.\\
  
\noindent \textbf{\large Born effective charges in undamped metals}  \\
In the presence of an electronic band gap far exceeding phonon characteristic frequencies, 
  the Born-Oppenheimer adiabatic approximation holds and $\bar{Z}_{s,\alpha\beta}(\omega,\Sigma^{\text{el}})\simeq \bar{Z}_{s,\alpha\beta}(0,0)$ in equation (\ref{eq:chivibr}) \cite{bistoni_giant_2019,PhysRevB.106.L180303}.
  Static BECs of semiconductors and insulators are routinely computed \textit{ab-initio} 
  using linear response techniques \cite{de_gironcoli_piezoelectric_1989,Gonze_prb1997,baroni_phonons_2001}.
  In metals with undamped electrons ($\Sigma^{\text{el}}=0$), the macroscopic dielectric response diverges at $(\mathbf{q}=0,\omega=0)$, $\bf{q}$ being the perturbation wavevector. To evaluate effective charges, we therefore need to either consider $(\mathbf{q}=0,\omega \neq 0)$ (dynamical case) or $(\mathbf{q}\neq0,\omega = 0)$ (static case), where the screening is finite. In the dynamical case, the polarization induced by time-dependent perturbations can be evaluated within a DFPT formulation, and one
  can define frequency-dependent BECs, $\bar{Z}_{s,\alpha\beta}^{\text{Dyn}}(\omega)$, 
  which are finite in the $\omega\rightarrow 0$ limit \cite{bistoni_giant_2019,binci_first-principles_2021,dreyer_nonadiabatic_2022,PhysRevB.106.L180303} (dynamical undamped BECs as a function of frequency are presented in \textbf{Extended Data Fig. 1}). %\ref{fig:dyn_becs}). 
  In the strictly static case, the requirement of zero macroscopic electric field at finite $\mathbf{q}$ fixes the $|{\bf q}|\rightarrow 0$  behaviour of the electrostatic potential. As a consequence, the macroscopic component of the screening of conducting electrons is removed, and we are allowed to define static BECs, $\bar{Z}_{s,\alpha\beta}^{\text{Stat}}$, from the ratio between the total charge density induced by a periodic atomic displacement and $|{\bf q}|$ \cite{macheda_2022,macheda2d}.
 In the undamped regime,  $\bar{Z}_{s,\alpha\beta}^{\text{Stat}}$ differs from $\bar{Z}_{s,\alpha\beta}^{\text{Dyn}}(\omega\rightarrow0)$ because of 
  an intraband contribution which is present in the static case but not in the dynamical one (Methods). To discuss such difference, 
  we use for the static approximation a slightly different but conceptually equivalent formulation to the one proposed in Refs.  \cite{macheda_2022,macheda2d}, leading to (Methods):
 \begin{align}
\bar{Z}_{s,\alpha\beta}^{\text{Stat}}=\bar{Z}_{s,\alpha\beta}^{\text{Dyn}}(i0^{+})+\Delta \bar{Z}_{s,\alpha\beta},
 \end{align}
where
 \begin{align}
    &\Delta \bar{Z}_{s,\alpha\beta} = +  \frac{2}{N_\mathbf{k}}\sum^{N_\mathbf{k}}_{\mathbf{k},l} f^{'}(\varepsilon_{\mathbf{k}}^{l})
    \bigg\langle u^{l}_{\mathbf k}\bigg|
     \frac{\partial \bar{V}^{\mathbf{\Gamma} }_{s\beta }}{i\partial q_{\alpha}}   
    \bigg|u^{l}_{\mathbf k}\bigg\rangle \nonumber \\
    &- \frac{2}{N_\mathbf{k}}\sum^{N_\mathbf{k}}_{\substack{\mathbf {k},lm\\m\ne l}}   f^{'}(\varepsilon_{\mathbf{k}}^{l})   
    \bigg\langle u^{l}_{\mathbf k}\bigg|{ \frac{i\hbar v^{\alpha}_{\mathbf{k}}}
    {\varepsilon_{\mathbf{k}}^{l} -\varepsilon_{\mathbf{k}}^{m} +i0^{+}}  }\bigg|{u^{m}_{\mathbf k} }\bigg\rangle 
    \bigg\langle u^{m}_{\mathbf k}\bigg| \bar{V}^{\mathbf{\Gamma} }_{s\beta } \bigg|u^{l}_{\mathbf k} \bigg\rangle .
    \label{eq_main:singul_step}
  \end{align}
Here $v^{\alpha}_{\mathbf{k}}=\frac{1}{\hbar}\frac{\partial H_{\mathbf k}}{\partial k^{\alpha}}$ is the electronic velocity operator, $f'(\varepsilon_{\mathbf{k}}^{l})$ is the derivative of the Fermi-Dirac distribution $f(\varepsilon_{\mathbf{k}}^{l})$ with respect to the electronic energy $\varepsilon_{\mathbf{k}}^{l}$ of band $l$ and crystal momentum $\mathbf k$, 
  $N_{\mathbf k}$ is the size of the momentum mesh and
  $\big|u^{l}_{\mathbf k}\rangle$ are the cell-periodic parts of the Bloch functions. $\bar{V}^{\mathbf{q} }_{s\beta }$ is the macroscopically unscreened (i.e., evaluated at zero macroscopic electric field) deformation potential as defined in equation (\ref{eq:def_pot}) for the displacement of the $s$ atom along the $\beta$ direction. 
The undamped dynamical BECs  have been recently shown to fulfill a sum rule relating them to the Drude weight, i.e., to the free-like portion of electron charge that is not bound to the underlying crystalline lattice \cite{dreyer_nonadiabatic_2022}. 
In the opposite static limit we derive (Methods) a sum rule, that reads: 
  \begin{align}
    \sum_s  \bar{Z}_{s,\alpha\beta}^{\text{Stat}} &= 
    n_{\varepsilon_{F}} \big\langle v^{\alpha}_{\mathbf{k}} p_{\mathbf k}^{\beta} \big\rangle_F
    -  n_{\varepsilon_{F}}  \sum_{ s} \bigg\langle\frac{\partial \bar{V}^{\mathbf{\Gamma} }_{s\beta }}{i\partial q_{\alpha}}\bigg\rangle_F,
    \label{eq_main:adiab_sumrule}
  \end{align}
  where $\mathbf{p}_{\mathbf k}=-i\hbar \nabla_{\mathbf r} + \hbar \mathbf k$ is the momentum operator,   $\big\langle{O_{\mathbf{k}}}\big\rangle_F=  - \frac{2}{N_\mathbf{k}}\sum^{N_\mathbf{k}}_{\mathbf{k},l}   f^{'}(\varepsilon_{\mathbf{k}}^{l})   
  \big\langle u^{l}_{\mathbf k}\big| O_{\mathbf{k}} \big| u^{l}_{\mathbf k} \big\rangle  /  n_{\varepsilon_{F}}$ defines the average over the Fermi surface and $n_{\varepsilon_{F}}=-\frac{2}{N_\mathbf{k}}\sum^{N_\mathbf{k}}_{\mathbf{k},l} f^{'}(\varepsilon_{\mathbf{k}}^{l})$ is the density of states at the Fermi level.
  It follows that the sum of BECs in the static limit is related to the kinetic energy averaged over the Fermi surface, plus an additional term that depends on the variation of the deformation potential. 
  Since both terms are proportional to $n_{\varepsilon_{F}}$, when applied to insulators one recovers the acoustic sum rule \cite{pick_microscopic_1970}, where BECs sum to zero. 
  In \ref{tab:h3s_al_sumrule} we summarize the numerical validation of the proposed sum rule for static BECs of H$_3$S and the conventional metal aluminium at different pressures. The static sum rule is respected with a precision better than 0.04\%  for all considered cases.
  
%Notice that,
We remark that the vibrational contribution to the
dielectric response should be computed at ${\bf q}={\bf 0}$ and at the finite frequencies $\omega=\omega_{\mu}$. Thus, in the undamped collisionless 
regime where $|\Sigma_{\text{el}}|\ll \hbar \omega_{\mu}$, the relevant BECs to be used in equation
(\ref{eq:chivibr}) are the dynamical ones, $\bar{Z}_{s,\alpha\beta}^{\text{Dyn}}(\omega_{\mu})$. This is the regime of the IR measurements performed in, e.g., neutral and doped bilayer graphene and in graphite, that are faithfully described by $\bar{Z}_{s,\alpha\beta}^{\text{Dyn}}(\omega_{\mu})$ \cite{PhysRevLett.103.116804,kuzmnenko_prl2008,Cappelluti_bigraphene_prb2012,binci_first-principles_2021,bistoni_giant_2019}.\\
  
\noindent \textbf{\large Born effective charges in overdamped and damped metals} \\
In presence of damping ($-\mathrm{Im}\Sigma^{\text{el}}\neq 0$), the divergences of the macroscopic dielectric response at $(\mathbf{q}=0,\omega=0)$ disappear and we can define BECs that display a well-defined continuous limit in such point.
In the Methods section we show that $\bar{Z}_{s,\alpha\beta}(\omega,\Sigma^{\text{el}})$ becomes equal to  $\bar{Z}_{s,\alpha\beta}^{\text{Stat}}$ in the collision-dominated (overdamped) regime, defined when $ \hbar \omega \ll  -\mathrm{Im} \Sigma^{\text{el}} \ll \hbar\omega_\mathrm{inter}$, i.e. when the scattering mechanism is far bigger than the frequency of the perturbation, but still doesn't relevantly affect interband transitions on their typical onset scale $\hbar\omega_\mathrm{inter}$. Indeed, in the overdamped regime the system relaxes instantaneously under the effect of an external perturbation, and therefore displays a static response (in the thermodynamical context, this is indicated as the `isothermal' response \cite{giuliani_vignale_2005}).
To address the intermediate damped regime  ($-\mathrm{Im} \Sigma^{\text{el}} \sim \hbar \omega$), we generalize to BECs the extended-Drude model (Methods), routinely employed for assessing the damped intraband contribution to optical conductivity, $\sigma_{\text{Drude}}(\omega,\Sigma^{\text{el}})$ \cite{Basov_RMP2005, Basov_RMP2011}, and already applied also to the study of nonadiabatic phonon frequencies  \cite{maksimov_nonadiabatic_1996}.
  %, to the case of BECs (see Methods Section). 
  This approach  accounts for finite-lifetime effects due to impurity and e-ph scattering within the Migdal-Eliashberg theory  \cite{marsiglio_eliashberg_2020} (Supplementary Section \ref{si:extdrude}). Eventually, adopting an isotropic approximation for the Brillouin zone integration (Methods), we find  the  expression for BECs in the regime of damped quasiparticles:
  \begin{align}
    \bar{Z}_{s,\alpha\beta}(\omega,\Sigma^{\text{el}}) &=
    \bar{Z}_{s,\alpha\beta}^{\text{Dyn}}(\omega) + \Delta \bar{Z}_{s,\alpha\beta}\, {I}(\omega,\Sigma^{\text{el}}),
    \label{eq:BECwithDrude0}
  \end{align}
  where [Methods, equation (\ref{eq:sigmadrude})]:
  \begin{align}
    {I}(\omega,\Sigma^{\text{el}})=
    \frac{\sigma_{\text{Drude}}(\omega,0) - \sigma_{\text{Drude}}(\omega,\Sigma^{\text{el}})}{\sigma_{\text{Drude}}(\omega,0) }.
    %=\frac{M(\omega,\Sigma^{\text{el}})}{\omega+M(\omega,\Sigma^{\text{el}})}.
    \label{eq:BECwithDrude1}
  \end{align}
Neglecting the frequency dependence of the self energy, $\Sigma^\text{el}\simeq i\mathrm{ Im}[\Sigma^\text{el}]= -i\Gamma$, results in the simpler expression:
  \begin{align}
    & \bar{Z}_{s,\alpha\beta}(\omega,-i\Gamma)  = 
    \bar{Z}_{s,\alpha\beta}^{\text{Dyn}}(\omega) + \Delta \bar{Z}_{s,\alpha\beta}
    \frac{i2\Gamma}{\omega+i2\Gamma}, 
    \label{eq:isotropicfactor}
  \end{align}
  highlighting the effect of quasiparticle damping onto BECs and recovering the dynamic and static expression in the undamped ($\Gamma\rightarrow 0$) and overdamped ($\Gamma\rightarrow \infty$) regimes, respectively. 
  
We emphasise that the overdamped regime is never attained 
in a superconductor for $\omega< 2\Delta$,  $\Delta$ being the SC gap,  because of the collisionless motion of superfluid condensate. Indeed, in the SC phase ${I}(0^{+},\Sigma^{\text{el}})=(1-x_{\rm SC})$, where $x_{\rm SC}$ is the condensate fraction, and thus $ \bar{Z}_{s,\alpha\beta}(0^+,\Sigma^{\text{el}})=x_{\rm sc}\bar{Z}_{s,\alpha\beta}^{\text{Dyn}}(i0^+)+(1-x_{\rm sc})\bar{Z}_{s,\alpha\beta}^{\text{Stat}}$.\\
  
\noindent \textbf{\large Infrared spectrum of H$_3$S superconductor } \\
We apply our theory to cubic H$_3$S at 150Gpa, that exhibits a SC critical temperature of about 200 K. 
We compute $\sigma_{\text{Drude}}(\omega,\Sigma^{\text{el}})$ and ${I}(\omega,\Sigma^{\text{el}})$ within a Migdal-Eliashberg approach for the normal ($T=$ 300 K) and superconducting ($T=$ 20 K) state, shown in \ref{fig:sigmaI}. This, combined with the DFT results for $\bar{Z}_{s,\alpha\beta}^{\text{Stat}}$ and $\bar{Z}_{s,\alpha\beta}^{\text{Dyn}}(\omega)$, allows us to obtain  BECs in the damped regime, $\bar{Z}_{s,\alpha\beta}(\omega,\Sigma^{\text{el}})$  that are displayed in \ref{fig:chargefig}. The difference between the static and dynamic limits ($\omega\rightarrow 0$) is remarkably large, of the same order of the BECs themselves, possibly leading to sign changes. The imaginary parts are also significantly affected in a broad range of frequencies, reflecting the marked frequency dependence and the complex nature of the dressing function $I(\omega,\Sigma^{\text{el}})$. Similar strong qualitative and quantitative differences in the real and imaginary parts of BECs are also visible when they are evaluated at the frequencies of optically active phonons (\ref{tab:h3s_dirty_bec_150-cubic}).

For phonon frequencies, displacements and lifetimes, we use the anharmonic results of Ref.  \cite{bianco_high-pressure_2018}. 
With these ingredients and $\bar{Z}_{s,\alpha\beta}(\omega,\Sigma^{\text{el}})$, we can compute the vibrational contribution of the dielectric response with equation (\ref{eq:chivibr}). The real part of corresponding optical conductivity, $\sigma_{\rm vib}(\omega)$, is presented in \ref{fig:sigmaI}\textbf{e},\textbf{f}. Interestingly, as remarked in \cite{PhysRevLett.103.116804,Cappelluti_bigraphene_prb2012,binci_first-principles_2021,bistoni_giant_2019},
the complex nature of the BECs results in a Fano resonance-profile in ${\rm Re}[\sigma_{\rm vib}(\omega)]$. Only in the superconducting phase, for $\omega < 2\Delta\simeq 85$ meV,  $\bar{Z}_{s,\alpha\beta}(\omega,\Sigma^{\text{el}})$ is real (neglecting the small resonant interband contribution of $\bar{Z}_{s,\alpha\beta}^{\text{Dyn}}(\omega)$, \ref{fig:chargefig}\textbf{bdf}) and the profile is Lorentzian, as in insulators.

Combining the electronic and vibrational conductivities, we can finally obtain the reflectivity spectrum, \ref{fig:R_excitations}\textbf{b}. The two IR phonon peaks are visible in both the normal and superconducting phase. The visibility of the higher-energy peak weakly depends on the impurity scattering rate $\eta_{\rm imp}$ (\ref{fig:R_vs_imp}). Instead, the lower-energy peak emerges only in presence of a strong impurity scattering, especially in the superconducting phase. This can be explained by the reduction with $\eta_{\rm imp}$ of the condensate fraction $x_{\rm SC}$ (\ref{fig:sigmaI}\textbf{g}), that dominates the electronic conductivity below, and near, $2\Delta$.
Albeit a quantitative comparison with experiments is hindered by the challenging requirements imposed by the extremely high pressure environment on reflectance measurements, requiring careful data processing to get rid of spurious contributions from diamond anvil cell and limiting the signal from the sample to a finite frequency range (110-190 meV) \cite{capitani_spectroscopic_2017}, the spectral position and profile of our simulated phonon features agree qualitatively well with experimental data. This confirms the identification of the drop experimentally observed around 150 meV in the IR spectrum of H$_3$S with a vibrational resonance of its cubic phase \cite{capitani_spectroscopic_2017}.\\

\noindent \textbf{\large Discussion and outlook } \\ 
Our theoretical formulation allows us to reconcile the two recently proposed definitions of BECs in metallic systems \cite{binci_first-principles_2021,bistoni_giant_2019,dreyer_nonadiabatic_2022,PhysRevB.106.L180303,macheda_2022,macheda2d}, at the same time providing a physical interpretation of the limiting regimes they describe. 
Accounting for finite lifetime effects of interacting quasiparticles onto BECs, and hence onto phonon oscillator strengths, enables the predictive simulation of vibrational optical spectra of bad conductors and metals with strong e-ph coupling.
Their interpretation represents an important tool for the structural characterization of hydrogen and superhydrides under extremely high pressures, for which reflectance measurements represent one of the few available experimental techniques \cite{capitani_spectroscopic_2017,Carbotte_prl2018,carbotte_spectroscopic_2019,Loubeyre_nature2020}. 

The proposed approach may find a natural application to other systems with strong e-ph coupling, as the broad class of metallic dichalcogenides where low-energy excitations in the IR range are optically accessible \cite{Wilson_TMDopt1969}, but typically analized within phenomenological Drude-Lorentz or Tauc-Lorentz models \cite{Dordevic_PRB2001,Gasparov_PRB2002,Munkhbat_acsphotonics2022} that use real BECs, neglect damping processes arising from the interaction with the electronic background. 
On the other hand, such damping processes are included in our approach via the same dressing function that models the effects of frequency-dependent scattering rates onto Drude-like contribution to the optical conductivity, suggesting its generalization to a much broader class of correlated materials \cite{Basov_RMP2011,IR_cuprates_review}. 

The emergence of new energy scales (as, e.g., the superconducting gap) and/or the presence of a strong collision regime leads to peculiar resonance effects also in Raman spectra of some strongly coupled superconductor \cite{Balseiro_prl1980,zeyher_zphysB1990}. Interestingly, such spectra have been interpreted in terms of damping processes of electronic excitations affecting the phonon spectral function, in both the superconducting \cite{zeyher_zphysB1990} and normal state \cite{Cappelluti_prb2006,saitta_giant_2008}, within a conceptual framework similar to that adopted here.

  \section*{Acknowledgments}\label{acknowledgments}
We thank L. Baldassarre, L. Benfatto, R. Bianco, J. Lorenzana, E. Nicol, and M. Ortolani for useful discussions and R. Bianco and E. Nicol  for providing us with supporting information from  \cite{bianco_high-pressure_2018,capitani_spectroscopic_2017}.
We acknowledge financial support from the European Union ERC-SYN MORE-TEM No 951215 (F. Mauri, G.M.), ERC DELIGHT No 101052708 (M.C.), Graphene Flagship Core3 No 881603 (F. Mauri, F. Macheda); and CINECA award under ISCRA initiative Grant no. HP10BV0TBS (P.B.) and HP10CSQ37Y  (G.M.) for the access to computational resources.
L. Binci acknowledges the Fellowship from the EPFL QSE Center ``Many-body neural simulations of quantum materials" (grant number 10060).
 Views and opinions expressed are however those of the author(s) only and do not necessarily reflect those of the European Union or the European Research Council. Neither the European Union nor the granting authority can be held responsible for them.

  \section*{Author Contributions Statement}\label{authors_contrib}
F. Mauri conceived the work and supervised the project with P. Barone.
All authors contributed to code developments in QUANTUM ESPRESSO and
EPIq \cite{marini_epiq_2023}.  G. Marchese performed all calculations with the support of F.
Macheda and L. Binci. G. Marchese, F. Macheda, P. Barone and F. Mauri
wrote the paper with contributions from M. Calandra and L. Binci.

  %\section*{Competing interests}\label{competing}
%The authors declare no competing interests.
  
  %\bibliography{h3s_prl}% Produces the bibliography via BibTeX.

  % \clearpage
\begin{figure*}[h]%
    \centering
    \includegraphics[width=0.7\textwidth]{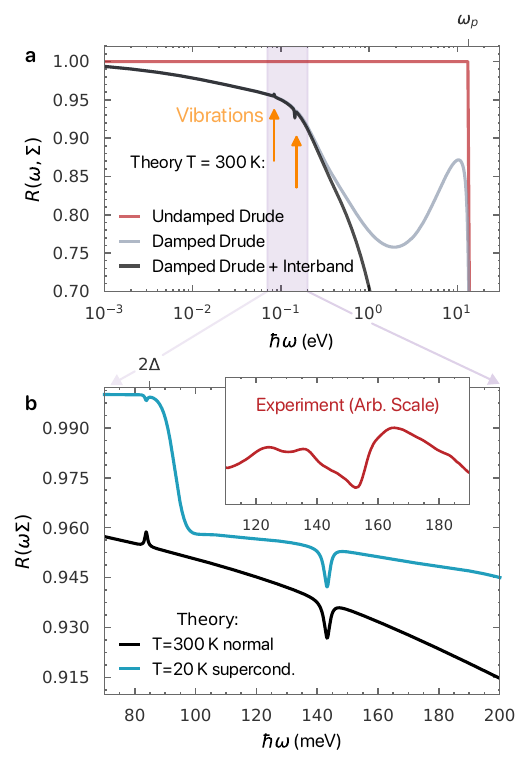}
    \caption{\textbf{Reflectivity spectrum for $\mathbf{H_3S}$ in the cubic \textit{Im$\bar{3}$m} phase at 150 GPa.} 
    ({\bf a}) Total \textit{ab-initio} reflectivity spectrum, including both electron and lattice contributions, evaluated at the experimental condition of $T= 300 $~K \cite{capitani_spectroscopic_2017} and shown up to the plasma frequency $\omega_p$. Different approximations are used to describe the electronic contribution: red and gray curves have been obtained neglecting interband contributions and assuming independent (`Undamped Drude', $\Sigma^{\text{el}}=0$) or interacting (`Damped Drude', $\Sigma^{\text{el}} \neq 0$) electrons. Vibrational resonances, indicated by vertical orange arrows, emerge only in the `Damped Drude' regime, being completely hidden in the `Undamped Drude' regime. Further inclusion of interband transitions (black curve) contributes at higher frequencies, thus not affecting the vibrational features.
    ({\bf b}) Closer view of the reflectivity in the IR range highlighted as a shadowed area in panel {\bf a}, for the normal phase at 300K and for the superconducting phase at 20K. Calculated vibrational features are compared with experimental data of Ref.  \cite{capitani_spectroscopic_2017} obtained at 300K, shown in the inset using an arbitrary scale. The spectral position of the measured drop agrees with our simulated spectrum, corroborating its assignment to an optically active phonon. In all the panels the electron self-energy $\Sigma^{\text{el}}$ is computed as explained in Supplementary Section \ref{si:extdrude} with an impurity scattering rate $\eta_{\text{imp}}$=135meV. The dependence of reflectivity on $\eta_{\text{imp}}$ is shown in \ref{fig:R_vs_imp}. }
    \label{fig:R_excitations}
  \end{figure*}

 \begin{figure*}[h]%
    \centering
    \includegraphics[width=0.9\textwidth]{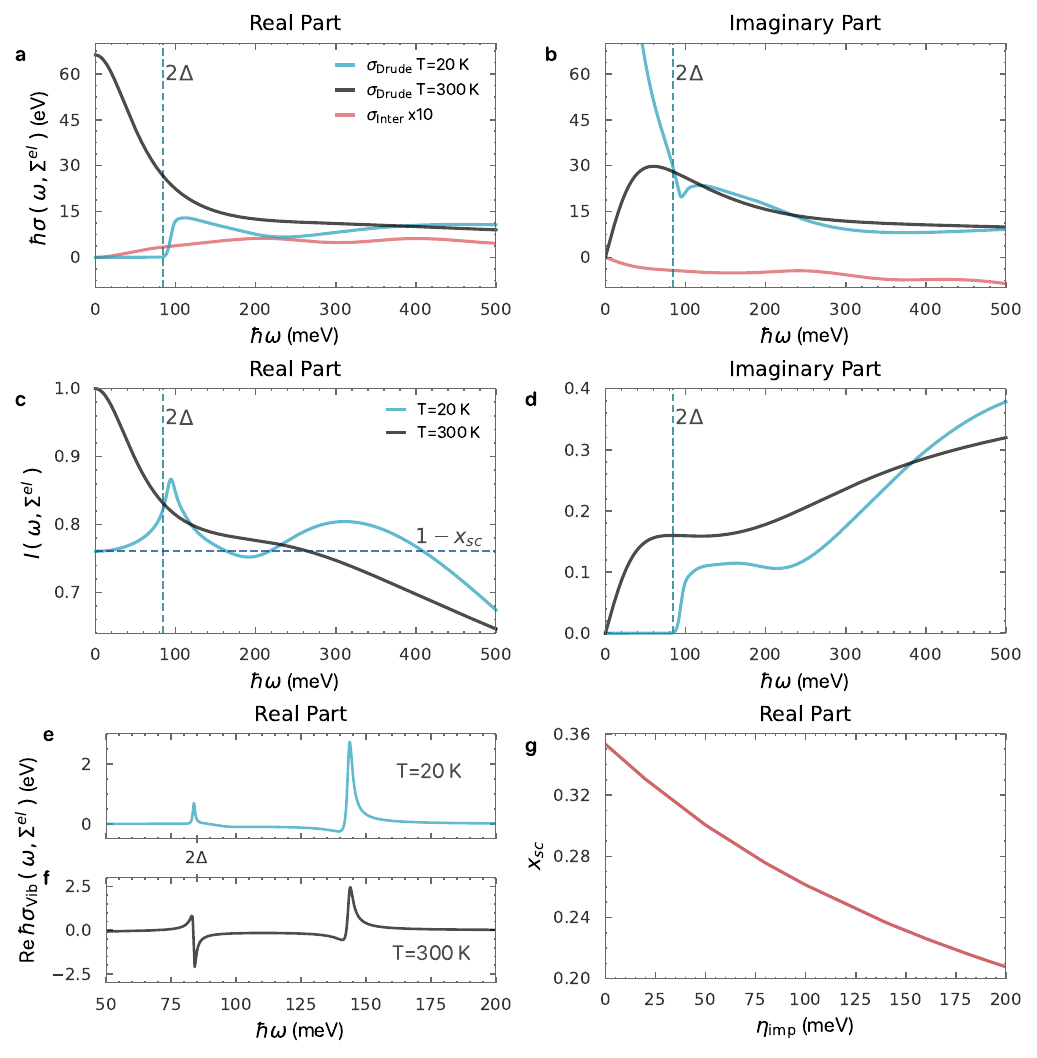} 
    \caption{\textbf{Electronic and vibrational optical conductivity.} 
    (\textbf{a}) Real and (\textbf{b}) imaginary parts of the electronic optical conductivity computed as explained in the Supplementary Section \ref{si:extdrude}, for the normal phase at 300 K (red curve) and for the superconducting phase at 20 K (blue curve), with an impurity scattering rate $\eta_{\text{imp}}$=135 meV. We present separately the Drude and the interband contribution, the latter magnified by a factor of ten being almost negligible in this energy range. From (\textbf{a},\textbf{b}) and Eq. (\ref{eq:sigmadrude}) we deduce (\textbf{c}) the real and (\textbf{d}) the imaginary parts of the adimensional factor $I(\omega,\Sigma^{\text{el}})$ defined in equation (\ref{eq:BECwithDrude1}). As discussed in Eq. (\ref{eq:static}) and in the main text, in the normal phase $I(0,\Sigma^{el})=1$ while in the superconducting phase $I(0^+,\Sigma^{el})=1-x_{\text{sc}}$. (\textbf{e,f}) Real part of the vibrational optical conductivity $\sigma_{\text{vib}}=-i\omega \chi_{\text{vib}}$ for the superconducting (\textbf{e}) and normal (\textbf{f}) phases. The non-Lorentzian shape of the peaks, due to the complex nature of BECs, is evident for the normal phase and for the phonon peak at 148 meV of the superconducting phase, while the peak at 84 meV recovers a Lorentzian shape since $I(\omega_1,\Sigma^{\text{el}})$ is real and $\mathrm{Im} Z^{\text{dyn}}(\omega_1)\sim 0$. (\textbf{g}) Behaviour of the condensate fraction, $x_{\rm SC}$, as a function of the impurity scattering rate, $\eta_{\text{imp}}$.}
    \label{fig:sigmaI}
  \end{figure*}
  
  \begin{figure*}[h]
    \centering
    \includegraphics[width=0.9\textwidth]{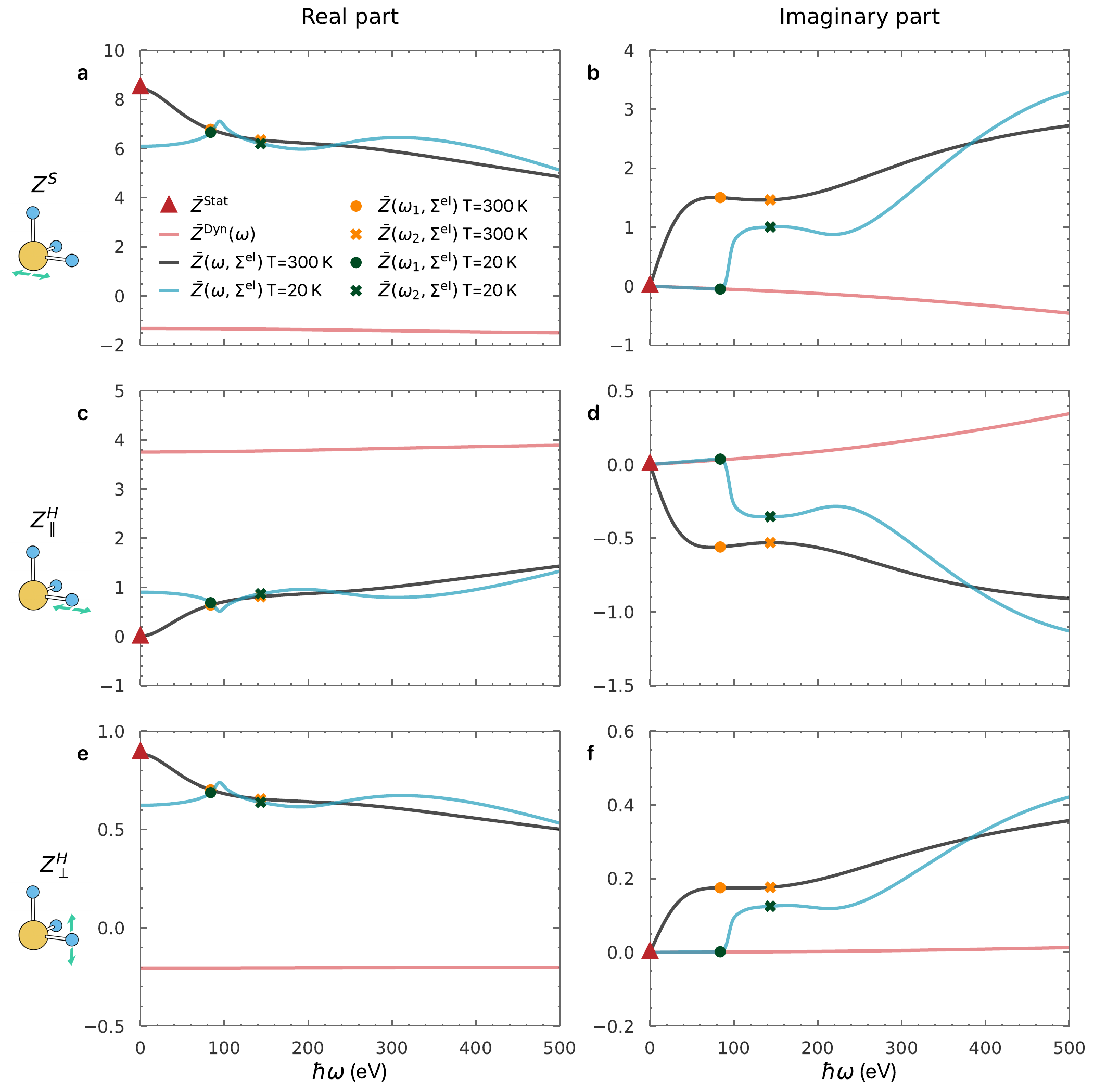}
    \caption{ \textbf{
      Born effective charges in the undamped, damped and overdamped regimes.} The real (\textbf{a},\textbf{c},\textbf{e}) and imaginary (\textbf{b},\textbf{d},\textbf{f}) parts of BECs in the dynamic undamped ($\bar{Z}_{s,\alpha\beta}^{\text{Dyn}}(\omega)$, red lines) and damped regimes ($\bar{Z}_{s,\alpha\beta}(\omega,\Sigma^{\rm el})$) for the normal  (black lines) and superconducting phases (blue lines) are shown as a function of frequency. Red triangles are the static, overdamped BECs ($\bar{Z}_{s,\alpha\beta}^{\text{Stat}}$). Orange and dark green dots highlight the BECs in the damped regime at the phonon frequencies $\hbar\omega_1=84~$meV and $\hbar\omega_2=148~$  meV, respectively, whose values are provided in  \ref{tab:h3s_dirty_bec_150-cubic}. Three independent components fully characterize BECs of the cubic $Im\bar{3}m$ phase of H$_3$S, as schematically shown in the lateral banner. Sulfur BEC tensor is fully isotropic and described by a single complex value (\textbf{a},\textbf{b}), while hydrogen BEC tensor is diagonal and comprises a longitudinal $Z_\parallel^H$ component (\textbf{c},\textbf{d}) and two transverse components $Z_\perp^H$ (\textbf{e},\textbf{f}) for H displacements parallel and perpendicular to the H-S bond, respectively.
      Undamped dynamical BECs acquire a more prominent frequency dependence at the onset scale of interband transitions (see \ref{fig:dyn_becs}) 
      }
      \label{fig:chargefig}
  \end{figure*}

  %\bibliographyMain{h3s_prl}
 
% \newpage
\clearpage
  % \appendix
  
  \noindent{\large \textbf{Methods}}

  \vspace{2ex}\noindent  {\bf Long-wavelength phonon perturbations in zero macroscopic electric field} \newline
  In Density Functional Theory (DFT) the electronic Hamiltonian is given by:
  \begin{align}
    H = \frac{\mathbf{p}^2}{2m}+V_{\text{KS}}=\frac{\mathbf{p}^2}{2m}+V_{\text{H}} + V_{\text{xc}} + V_{\text{ext}},
  \end{align}
  where $V_{\text{ext}}$ is the electron-ion interaction, $V_{\text{H}} $ is the Hartree term and $V_{\text{xc}} $ is the exchange-correlation potential. The atoms are positioned in ${\bf u}^{\mathbf{R}}_{s}=\mathbf R+\bm{\tau}_s $, $\mathbf{R}$ being a Bravais  vector and $\bm{\tau}_s$ an internal coordinate.  Collective displacements with wavevector $\mathbf{q}$ of the atoms $s$ along the Cartesian axes $\beta$ induce a cell-periodic potential variation $V^{\mathbf{q}}_{s\beta}(\mathbf{r})$ defined by:
  \begin{align}
    e^{i{\mathbf q}\cdot \mathbf{r}} V^{\mathbf{q}}_{s\beta}(\mathbf{r})  &= \frac{\partial V_{\text{KS}}(\mathbf r)}{ \partial  u^{\mathbf{q}}_{s\beta}} =\sum_{\mathbf R} e^{i{\mathbf q}\cdot( \mathbf{R} +\bm{\tau}_s  ) }  \frac{\partial  V_{\text{KS}}(\mathbf r)}{ \partial  u^{\mathbf{R}}_{s\beta}} .
  \end{align}
At zone center, $V^{\boldsymbol{\Gamma}}_{s\beta}$ should not contain the divergent part of the transform of the Coulomb potential in reciprocal space \cite{Gonze_prb1997}; therefore, in metals, it is corrected introducing a `Fermi level ($\varepsilon_{F}$) shift' $F^S_{s\beta}={\partial \varepsilon_{F}}/{\partial u^{\boldsymbol{\Gamma}}_{s\beta}}$ to enforce charge neutrality \cite{baroni_phonons_2001}. We expand the above treatment of the Coulomb divergence and of the Fermi level shift, from the single $\mathbf{\Gamma}$ point -- as routinely done within Density Functional Perturbation Theory (DFPT)  \cite{baroni_phonons_2001}~-- to a small neighborhood around it. This allows us to introduce a macroscopically unscreened deformation potential $\bar{V}^{\mathbf q}_{s\beta}(\mathbf{r})$, continuous in the origin, that induces BECs free from the influence of macroscopic electronic screening (for more details, see Refs.  \cite{macheda_2022,macheda2d,pick_microscopic_1970} \cite{PhysRevX.11.041027,PhysRevB.88.174106}). Within the typical case and notation of semiconductors, as shown in Refs.  \cite{macheda_2022}\cite{PhysRevX.11.041027}, such a procedure leads to the computation of $Z^*_{s,\alpha\beta}$ and not of $Z^*_{s,\alpha\beta}/\epsilon_{\infty}$, where $\epsilon_{\infty}$ is the high-frequency dielectric constant, and can thus be extended also to the metallic case where the screening is divergent. We therefore enforce two conditions in a small sphere of radius $\mathcal{Q}$ centered at $\mathbf{\Gamma}$ and contained in the first Brillouin zone: (i) null macroscopic electric field as by definition of BECs  \cite{baroni_phonons_2001}, (ii) absence of monopolar charge terms per unit volume. Coherently with these two requests, 
%the Fourier components of $\bar{V}^{\mathbf q}_{s\beta}(\mathbf{r})$ are
we introduce in reciprocal space representation
  %
  %\begin{align}
    %\bar{V}^{\mathbf q}_{s\beta}(\mathbf{r})=\left(V_{\text{xc}}\right)^{\mathbf{q}}_{s\beta}(\mathbf{r})+ 
    %\begin{cases}
      %-F^{\text{S}}_{s\beta} 
      %&|{\bf q}| < \mathcal{Q} \\
      %\left({V}_\text{extH}\right)^{\mathbf q}_{s\beta}(\mathbf{r})  & |{\bf q}|  > \mathcal{Q}  \\
    %\end{cases},
    %\label{eq:def_pot}
  %\end{align}
%
  %\begin{align}
    %\bar{V}^{\mathbf q}_{s\beta}(\mathbf{G})=\left(V_{\text{xc}}\right)^{\mathbf{q}}_{s\beta}(\mathbf{G})+ 
    %\begin{cases}
      %-F^{\text{S}}_{s\beta} 
      %&|{\bf q}+ {\bf G}| < \mathcal{Q} \\
      %\left({V}_\text{extH}\right)^{\mathbf q}_{s\beta}(\mathbf{G})  & |{\bf q}+ {\bf G}|  > \mathcal{Q}  \\
    %\end{cases},
    %\label{eq:def_pot}
  %\end{align}
  \begin{align}
    \bar{V}^{\mathbf q}_{s\beta}(\mathbf{G})=\left(V_{\text{xc}}\right)^{\mathbf{q}}_{s\beta}(\mathbf{G})+ 
    \begin{cases}
      -F^{\text{S}}_{s\beta} 
      &|{\bf q}| < \mathcal{Q} \text{, } G=0\\
      \left({V}_\text{extH}\right)^{\mathbf q}_{s\beta}(\mathbf{G})  & \text{ otherwise}  \\
    \end{cases},
    \label{eq:def_pot}
  \end{align}
  where we have called $V_{\text{ext}\text H} = V_{\text{ext}} + V_{\text H}$, $\mathbf{G}$ are reciprocal lattice vectors
  %, $\mathbf{q}$ belongs to first Brillouin zone 
  and the value of $F^{\text{S}}_{s\beta}$ is chosen  so as  $\lim_{\mathbf{q}\rightarrow 0 } \left(\bar{\rho}^{\text{tot}}\right)^{\mathbf{q}}_{s\beta}=  0$, where $\left(\bar{\rho}^{\text{tot}}\right)^{\mathbf q}_{s\beta}$ is the total density (electronic plus ionic)  induced by $\bar{V}^{\mathbf q}_{s\beta}$. This last requirement corresponds to the condition
  \begin{align}
    \frac{2}{N_\mathbf{k}}\sum^{N_\mathbf{k}}_{\mathbf{k},l} f^{'}(\varepsilon_{\mathbf{k}}^{l}) 
    \big\langle u^{l}_{\mathbf k}\big| \bar{V}^{\mathbf{\Gamma} }_{s\beta }\big| u^{l}_{\mathbf k}\big\rangle = 0.\label{eq:FermiS}
  \end{align}
  At $|{\bf q}|=0$, equations (\ref{eq:def_pot},\ref{eq:FermiS}) are equivalent to equations (38-43) of   \cite{calandra_adiabatic_2010} and equation (79) of  \cite{baroni_phonons_2001}. At $0<|{\bf q}|<\mathcal{Q}$, if $F^{\text{S}}_{s\beta}=0$ by symmetry,  the present unscreening procedure is fully equivalent, within the 
  Random Phase Approximation, to that of Refs. \cite{macheda_2022,macheda2d}. In the same regime, if $F^{\text{S}}_{s\beta}\ne 0$, the equivalence holds if the effects of local fields on the response are neglected.  Finally note that in the two metals treated in the present work, H$_3$S and elemental Al, $F^{\text{S}}_{s\beta}= 0$ by symmetry.

    \vspace{2ex}\noindent  \textbf{Unscreened static BECs} \newline
    As extensively discussed in Refs.  \cite{macheda_2022,macheda2d}, static BECs are the sum of an ionic and an electronic term, i.e. $\bar Z^{\text{Stat.}}_{s,\alpha \beta}=\bar Z^{\text{ion}}_{s,\alpha \beta}+\bar Z^{\text{el}}_{s,\alpha \beta} $. The ionic contribution amounts to the valence charge number $\bar Z^{\text{ion}}_{s,\alpha \beta}=Z^{\text{ion}}_{s}\delta_{\alpha \beta}$, while the electronic term can be derived from macroscopic electrostatics. In the typical linear response problem, the electronic charge $\left(\rho^{\text{el}}\right)_{s\beta}^{\mathbf{q}}(\mathbf{G=0})=\rho_{s\beta}(\mathbf{q})$ induced by $u^{\mathbf{q}}_{s\beta}$ determines a macroscopic electronic polarization $\left(\mathbf{P}^{\text{el}}\right)_{s\beta}(\mathbf{q})=\partial\mathbf{P}^{\text{el}}/\partial u^{\mathbf{q}}_{s\beta}$ given by
  \begin{align}
    i\mathbf{q}\cdot \left(\mathbf{P}^{\text{el}}\right)_{s\beta}(\mathbf{q})=
    -\rho_{s\beta}\left( \mathbf{q}\right).
  \end{align}
  If instead the linear response problem is modified according to equation (\ref{eq:def_pot}), taking ${\mathbf q}=q \hat{\bm{\alpha}}$ we can define
  \begin{align}
    \frac{- e}{i\Omega} {\bar Z}^{\text{el}}_{s,\alpha \beta}  = i  \left( \bar {P}^{\text{el}}_{\alpha}\right)_{s\beta}(\mathbf{0}) 
    = -\lim_{q\rightarrow 0} \frac{\bar{\rho}_{s\beta}\left( q\bm{\hat{\alpha}} \right)}{q};
    \label{eq:defzbar}
  \end{align}
  the limit of the above expression is finite in both metals and insulators. In particular, the induced density may be written, starting from the expressions of Ref.  \cite{baroni_phonons_2001}, as
  \begin{align}
    \bar{\rho}_{s\beta}(\mathbf q) = 
    \frac{-e}{\Omega}\frac{2}{N_\mathbf{k}}\sum^{N_\mathbf{k}}_{\mathbf{k},lm} &
    \frac{f(\varepsilon^{l}_{\mathbf k})-f(\varepsilon^{m}_{\mathbf k+\mathbf q})}
    {\varepsilon^{l}_{\mathbf k}-\varepsilon^{m}_{\mathbf k+\mathbf q}} 
    \times \nonumber\\&
    \big\langle u^{l}_{\mathbf k}\big|u^{m}_{\mathbf k + \mathbf q}\big\rangle  
    \big\langle u^{m}_{\mathbf k + \mathbf q}\big| \bar{V}^{\mathbf{q} }_{s\beta }\big| u^{l}_{\mathbf k}\big\rangle,
    \label{eq:bubblerho}
  \end{align}
  which manifestly shows a linear behaviour in the limit $\mathbf{q}\rightarrow 0$ (as explicitly shown below), and therefore leads to a finite value for the static BECs. In insulators, $\bar Z^{\text{Stat.}}_{s,\alpha \beta}$  coincides with $Z^*_{s,\alpha\beta}$.
    
    \vspace{2ex}\noindent  \textbf{Explicit expression for static BECs} \newline
     In order to obtain the long-wavelength limit of $\bar{\rho}_{s\beta}(\mathbf q)$, we distinguish between the contribution of the electron-hole propagator, $B_{\mathbf{k},lm}(\mathbf q)$ and of the matrix elements, $M^{s\beta}_{\mathbf{k},lm}(\mathbf q)$, writing
  \begin{align}
    \bar{\rho}_{s\beta}(\mathbf q) &= 
    \frac{e}{i\Omega}\frac{2}{N_\mathbf{k}}\sum^{N_\mathbf{k}}_{\mathbf{k},lm} 
    B_{\mathbf{k},lm}(\mathbf q) M^{s\beta}_{\mathbf{k},lm}(\mathbf q).
    \label{eq:rho_factorized}
  \end{align}
  Interpreting $\bar{\rho}_{s\beta}(\mathbf q)$ as a two-points correlation function, we express the bubble part as 
  \begin{align}
    B_{\mathbf{k},lm}(\mathbf q) &=
    \frac{f(\varepsilon^{l}_{\mathbf k})-f(\varepsilon^{m}_{\mathbf k+\mathbf q})}
    {\varepsilon^{l}_{\mathbf k}-\varepsilon^{m}_{\mathbf k+\mathbf q}} \nonumber \\
    &=- k_B T \sum_{i\omega_n}g^{m}_{\mathbf{k}+\mathbf{q}}(i\omega_n)g^{l}_{\mathbf{k}}(i\omega_n)
    \label{eq:stat_bubble}
  \end{align}
  where  $g^l_\mathbf{k}(z)=\frac {1}{\hbar z-\varepsilon^{l}_{\mathbf k}}$ is the non-interacting  electron Green function, $i\omega_n$ are the Fermionic Matsubara frequencies.
  And the vertices are
  \begin{align}
    M^{s\beta}_{\mathbf{k},lm}(\mathbf q) &= -i \big\langle u^{l}_{\mathbf k}\big|u^{m}_{\mathbf k + \mathbf q}\big\rangle 
    \big\langle u^{m}_{\mathbf k + \mathbf q}\big|\bar{V}^{\mathbf{q} }_{s\beta }\big|u^{l}_{\mathbf k}\big\rangle.
    \label{eq:vertices}
  \end{align}
  We can expand $B$ and $M$ in powers of the momentum $q$:
  \begin{align}
    B_{\mathbf {k},lm}(q \hat{\bm{\alpha}}) &=  B^0_{\mathbf {k},lm} 
    + q \,B^{1,\alpha}_{\mathbf {k},lm} + \mathcal{O}(q^2),
    \nonumber \\
    M^{s\beta}_{\mathbf {k},lm}(q \hat{\bm{\alpha}}) &=  M^{0,s\beta}_{\mathbf {k},lm} 
    + q \,M^{1,s\alpha\beta}_{\mathbf {k},lm} 
    + \mathcal{O}(q^2).
  \end{align}
  with
  \begin{align}
    M^{1,s\alpha\beta}_{\mathbf {k},lm}=&\bigg\langle u^{l}_{\mathbf k}\bigg| 
    \left.
    \frac{\partial|u^{m}_{{\mathbf k}+{\mathbf q}}\rangle 
    \langle u^{m}_{{\mathbf k}+{\mathbf q}}|}{i\partial q_{\alpha}}\right\vert_{{\mathbf q}={\bf 0}}
    {\bar{V}^{\mathbf{\Gamma} }_{s\beta }}\bigg|u^{l}_{\mathbf k}\bigg\rangle \nonumber \\
    &+ \delta_{lm}\bigg\langle{u^{l}_{\mathbf k }}\bigg|
    {\frac{\partial \bar{V}^{\mathbf{\Gamma} }_{s\beta } }{i\partial q_{\alpha}}}\bigg|{u^{l}_{\mathbf k}}\bigg\rangle.
  \end{align}
  Using $\varepsilon^{m}_{\mathbf k+\mathbf q} = \varepsilon^{m}_{\mathbf{k}} + q \big\langle u^{m}_{\mathbf k}\big|  \hbar  v^{\alpha}_{\mathbf{k}} \big|u^{m}_{\mathbf k} \big\rangle
  + \mathcal{O}(q^2) $ and  $\mathbf{k}\cdot\mathbf{p}$ perturbation theory to evaluate $|u^{m}_{\mathbf k+ \mathbf q}\rangle$, we obtain for the intraband ($l=m$)  case: 
  \begin{align}
    B^0_{\mathbf{k},ll}  &=  f'(\varepsilon_{\mathbf{k}}^{l}) 
    \;,
    \qquad
    M^{0,s\beta}_{\mathbf{k},ll}  = -i \big\langle u^{l}_{\mathbf k} \big| \bar{V}^{\mathbf{\Gamma} }_{s\beta }\big|u^{l}_{\mathbf k}\big\rangle,
    \nonumber \\
    B^{1,\alpha}_{\mathbf {k},ll}  &= \frac{1}{2} f''(\varepsilon_{\mathbf{k}}^{l})
    \big\langle u^{l}_{\mathbf k}\big|  \hbar  v^{\alpha}_{\mathbf{k}} \big|u^{l}_{\mathbf k}\big\rangle ,
    \nonumber \\
    M^{1,s\alpha\beta}_{\mathbf {k},ll}  &= 
    - \sum_{m',m'\ne l}   
    \bigg\langle{u^{l}_{\mathbf k}}\bigg|{  \frac{i \hbar v^{\alpha}_{\mathbf{k}}}{\varepsilon^{l}_{\mathbf k} - \varepsilon^{m'}_{\mathbf k}}    }\bigg|{u^{m'}_{\mathbf k}}\bigg\rangle
    \big\langle{u^{m'}_{\mathbf k}}\big|{\bar{V}^{\mathbf{\Gamma} }_{s\beta }}\big|{u^{l}_{\mathbf k}}\big\rangle
    \nonumber\\ &\qquad
    + \bigg\langle{u^{l}_{\mathbf k }}\bigg|
    {\frac{\partial \bar{V}^{\mathbf{\Gamma} }_{s\beta } }{i\partial q_{\alpha}}}\bigg|{u^{l}_{\mathbf k}}\bigg\rangle,
    \label{M1dia}
  \end{align}
  and for the interband ($l\ne m$) case:
  \begin{align}
    B^0_{\mathbf{k},lm,m \ne l}  &= \frac{f(\varepsilon^{l}_{\mathbf k})-f(\varepsilon^{m}_{\mathbf k})}{\varepsilon_{\mathbf{k}}^{l} -\varepsilon_{\mathbf{k}}^{m} +i0^+} \;,
    \qquad  M^{0,s\beta}_{\mathbf{k},lm,m \ne l}  =  0,
    \nonumber \\
    B^{1,\alpha}_{\mathbf {k},lm,m \ne l}  &= \frac{
    \big\langle u^{m}_{\mathbf k}\big|\hbar v^{\alpha}_{\mathbf{k}}\big|u^{m}_{\mathbf k}\big\rangle }
    {\varepsilon_{\mathbf{k}}^{l} -\varepsilon_{\mathbf{k}}^{m}}  
    {\left[\frac{f(\varepsilon^{l}_{\mathbf k})-f(\varepsilon^{m}_{\mathbf k})}{\varepsilon_{\mathbf{k}}^{l} -\varepsilon_{\mathbf{k}}^{m}} - f'(\varepsilon^{m}_{\mathbf{k}})  
    \right]},   
    \nonumber \\
    M^{1,s\alpha\beta}_{\mathbf {k},lm,m \ne l}  &= 
    -\bigg\langle{u^{l}_{\mathbf k}}\bigg|{  \frac{i\hbar v^{\alpha}_{\mathbf{k}}}{\varepsilon^{m}_{\mathbf k} - \varepsilon^l_{\mathbf k}}    }\bigg|{u^{m}_{\mathbf k}}\bigg\rangle
    \big\langle u^{m}_{\mathbf k}\big|\bar{V}^{\mathbf{\Gamma} }_{s\beta }\big|u^{l}_{\mathbf k}\big\rangle .
  \end{align}
  Due to the condition equation (\ref{eq:FermiS}), the term  
  $\sum^{N_\mathbf{k}}_{\mathbf{k},l}
  B^0_{\mathbf {k},ll} M^{0,s\beta}_{\mathbf {k},ll}$ is zero.
  In addition, in systems with Time Reversal Symmetry (TRS) the following identities hold:   
  $\big\langle u^{m}_{-\mathbf k}\big|\bar{V}^{\mathbf{\Gamma} }_{s\beta }\big|u^{l}_{-\mathbf k}\big\rangle  = \big\langle u^{l}_{\mathbf k}\big|\bar{V}^{\mathbf{\Gamma} }_{s\beta }\big|u^{m}_{\mathbf k}\big\rangle $ and $
  \big\langle u^{m}_{-\mathbf k}\big|v^{\alpha}_{\mathbf{-k}}\big|u^{l}_{-\mathbf k}\big\rangle = -\big\langle u^{l}_{\mathbf k}\big|v^{\alpha}_{\mathbf{k}}\big|u^{m}_{\mathbf k}\big\rangle .
  $
  Thus, we find out that the term $B^{1,\alpha}_{\mathbf {k},ll}M^{0,s\beta}_{\mathbf {k},ll}$ is odd under TRS and disappears in the integration over the Brillouin zone. Hence, inserting the induced density at $\mathcal{O}(q^2)$ in equation (\ref{eq:defzbar}) we find
  \begin{align}
    \bar{Z}^{\text{Stat}}_{ s,\alpha \beta} &= \bar Z^{\text{Dyn}}_{s,\alpha\beta}(i0^+)
    +\Delta     \bar{Z}_{s,\alpha\beta}, 
    \label{eq:bec_at_1st}
  \end{align}
  where $\bar Z^{\text{Dyn}}_{s,\alpha\beta}(i0^+)$, coherently with Refs.  \cite{binci_first-principles_2021,dreyer_nonadiabatic_2022}, is equal to  
  \begin{align}
    \bar{Z}^{\text{Dyn}}_{s,\alpha\beta}(i0^+) &=  
    Z^{\text{ion}}_{s} \delta_{\alpha\beta} +      \frac{2}{N_\mathbf{k}} \sum^{N_\mathbf{k}}_{\substack{\mathbf{k},lm\\m\ne l}}
    B^0_{\mathbf {k},lm}  M^{1,s\alpha\beta}_{\mathbf {k},lm}
    \label{eq:dyn_bec}
  \end{align}
   and $  \Delta \bar{Z}_{s,\alpha\beta}$, defined in equation (\ref{eq_main:singul_step}), can be expressed as
  \begin{align}
    \Delta \bar{Z}_{s,\alpha\beta} &=
    \frac{2}{N_\mathbf{k}}
    \sum^{N_\mathbf{k}}_{\mathbf{k},l}
    B^0_{{\mathbf {k}},ll}  M^{1,s\alpha\beta}_{{\mathbf {k}},ll}
    \label{eq:deltaZ_as_prod}.
  \end{align}
  Notice that the static limit is well defined by equation (\ref{eq:defzbar}) so that in principle the imaginary part $i0^+$ is unnecessary. Such regularization is required only when separating individual divergences whose sum cancels out.
  % However, to compare with the dynamic limit, which inherits intrinsically such an imaginary regularization, we explicit its presence.
We stress its presence here in view of comparing the static and the dynamic limit, which inherits such imaginary regularization from causality.

   \vspace{2ex}\noindent  \textbf{Static BECs sumrule} \newline
  \noindent  The BECs sumrule is based on the equivalence between acoustic modes and translations as expressed by the following identity
  \begin{align}
    \sum_{\mathbf{s}} \bar{V}^{\mathbf{\Gamma} }_{s\beta }  =\sum_{\mathbf{R},s} \frac{\partial  V_{\text{KS}}}{\partial  u^{\mathbf{R}}_{s\beta}}=\frac{1}{i\hbar}[p^{\beta},V_{	\text{KS}}]=\frac{1}{i\hbar}[p_{\mathbf{k}}^{\beta},H_{\mathbf{k}}],
  \end{align}
  from which we derive
  \begin{align}
    \sum_{s} \big\langle u^{m}_{\mathbf{k}}\big|\bar{V}^{\mathbf{\Gamma} }_{s\beta }\big|u^{l}_{\mathbf{k}}\big\rangle=   
    \bigg\langle{u^{m}_{\mathbf{k}}}\bigg|
    {\frac{p_{\mathbf{k}}^{\beta}}{i\hbar}  }\bigg|{u^{l}_{\mathbf{k}}} \bigg\rangle
    ( \varepsilon_{\mathbf{k}}^{l} -\varepsilon_{\mathbf{k}}^{m}).
    \label{eq:g2p}
  \end{align}
  As described in the Supplementary Section \ref{si:becs_sumrule}, using equation (\ref{eq:g2p}), we find
  \begin{align}
   \sum_s\Delta \bar{Z}_{s,\alpha\beta} & = \sum_s \frac{2}{N_\mathbf{k}}\sum^{N_\mathbf{k}}_{\mathbf{k},l} B^{0}_{\mathbf{k},ll} M^{1,s\alpha\beta}_{\mathbf{k},ll}
    = n_{\varepsilon_F}\big\langle v^{\alpha}_{\mathbf{k}} p^{\beta}_{\mathbf{k}}\big\rangle_F 
    \nonumber\\
    -   \frac{m \Omega}{4 \pi e^2} (\omega_{\text{p}}^2)_{\alpha\beta} 
    &-  n_{\varepsilon_{F}}  \sum^{N_{\text{el}}}_{ s} \bigg\langle\frac{\partial \bar{V}^{\mathbf{\Gamma} }_{s\beta }}{i\partial q_{\alpha}}\bigg\rangle_F,
    \label{eq:delta_z_summed}
  \end{align}
  where the plasma frequency is defined  as  (see also Supplementary Section \ref{si:vel_op}):
  \begin{align}
    (\omega_{\text{p}}^2)_{\alpha\beta}= -  \frac{4 \pi e^2}{\Omega}\frac{2}{N_\mathbf{k}}\sum^{N_\mathbf{k}}_{\mathbf{k},l}   f^{'}(\varepsilon_{\mathbf{k}}^{l})   
    \big\langle u^{l}_{\mathbf k}\big| v^{\alpha}_{\mathbf k }\big| u^{l}_{\mathbf k }\big\rangle 
    \bigg\langle{u^{l}_{\mathbf k }}\bigg|{\frac{p^{\beta }_{\mathbf k }}{m}}\bigg|{u^{l}_{\mathbf k}}\bigg\rangle.%\nonumber
  \end{align}
  By combining equation (\ref{eq:delta_z_summed}) with equation (6) in Ref.  \cite{dreyer_nonadiabatic_2022}, namely $\sum_s \bar{Z}^{\rm Dyn}_{s,\alpha\beta}(i0^+)=\frac{m \Omega}{4 \pi e^2} (\omega_{\text{p}}^2)_{\alpha\beta}$,  we obtain  the sumrule for static BECs, equation (\ref{eq_main:adiab_sumrule}). We validate this sumrule in H$_3$S and Al at different pressures and reported the results  in \ref{tab:h3s_al_sumrule}. 

    \vspace{2ex}\noindent  \textbf{BECs in the damped regime} \newline
  \noindent  The evaluation of BECs in the damped regime can be performed in certain approximations. Exploiting an anisotropic `dressed-bubble'  \cite[p. 507, equation 8.45]{mahan_many-particle_2000} approach, we find an expression that reduces to the dynamic and static BECs in the undamped and overdamped limits, as physically expected  \cite{pines_theory_1989}. 

  We assume that the effect of the scattering upon interband ($l\ne m$) contribution is negligible, i.e. $-\mathrm{Im} \Sigma^{\text{el}} \ll \hbar \omega_{inter} $. Hence, we approximate the interband contribution with its clean value while dressing the intraband term. As for the static BECs in the clean regime, we expand the vertices and the electron-hole propagator up to linear order in $\bf q$. TRS still implies that the first order in $\mathbf{q}$ of the bubble ($B^1$) does not contribute. Furthermore, we neglect vertex corrections.
  For $B^0$ we write
\begin{align}
  &\scalebox{.925}{$\displaystyle
    B^{0}_{\mathbf{k},ll}(\omega,\Sigma^{\text{el}}) = \left.- k_B T \sum_{i\omega_n}G^{l}_{\mathbf{k}}(i\omega_n)G^{l}_{\mathbf{k}}(i\omega_n+i\omega_p)\right\vert_{i\omega_{\text{p}}\Rightarrow \hbar \omega+ i0^+}$
    }
    \nonumber\\    
    &\scalebox{.95}{$\displaystyle=   
    \int_{\mathbb{R}^2} dx dy
    \mathcal{A}^{l}_{\mathbf{k}}(x)\mathcal{A}^{l}_{\mathbf{k}}(y)
    \frac{f(\varepsilon^{l}_{\mathbf{k}}+x) - f(\varepsilon^{l}_{\mathbf{k} }+y)}
    {(\varepsilon^{l}_{\mathbf{k}}+ x) - (\varepsilon^{l}_{\mathbf{k} }+y) + \hbar \omega + i0^{+}}
    $}
    \label{eq:rho_intra_dress},
  \end{align}  
  where  $i\omega_{\text{p}}$  are the Bosonic Matsubara frequencies, $\Rightarrow$ indicates the analytic continuation, $G^{l}_{\mathbf{k}}(z) = \int_{\mathbb{R}} dx \frac{\mathcal{A}^{l}_{\mathbf{k}}(x)}{\hbar z-  \varepsilon^{l}_{\mathbf{k}} -x } $ is the interacting  Green function,
  \begin{align}
    \mathcal{A}^{l}_{\mathbf{k}}(x)=-\frac{1}{\pi}\mathrm{ Im}\left[\frac{1}{x+i0^+-\Sigma^{\text{el}}_{\mathbf{k}l}(x+i0^+ +\varepsilon^{l}_{\mathbf{k}})}\right]
  \end{align}
  is the spectral function centered at the non-interacting energy $\varepsilon^{l}_{\mathbf{k}}$, with $\mathcal{A}^{l}_{\mathbf{k}}(x)\ge 0$ and $\int_{\mathbb{R}}dx \mathcal{A}^{l}_{\mathbf{k}}(x)=1$  \cite{mahan_many-particle_2000}.

  A further assumption is that the induced density weakly depends on the electronic temperature used in the Fermi-Dirac function of equation~(\ref{eq:rho_intra_dress}). This assumption is respected in many metallic systems (and in particular in H$_3$S) since the electronic bands energies vary on the scale of the eV. We can then, within a good approximation, choose a temperature that satisfies $k_B T\gg \lvert \Sigma^{\text{el}}_{\mathbf{k}}\rvert$ and expand the electronic occupation as
  \begin{align}
    f(\varepsilon^{l}_{\mathbf{k}} + y) \simeq f(\varepsilon^{l}_{\mathbf{k}}) + y f'(\varepsilon^{l}_{\mathbf{k}}),
  \end{align}
  to obtain
  \begin{align}
    B^0_{\mathbf{k},ll}(\omega,\Sigma^{\text{el}}) &\simeq
    f'(\varepsilon^{l}_{\mathbf{k}}) I^{l}_{\mathbf{k}}(\omega,\Sigma^{\text{el}}),  
    \label{eq:spectral_bubble}
  \end{align}
  where we introduced the adimensional function
  \begin{align}
    I^{l}_{\mathbf{k}}(\omega,\Sigma^{\text{el}}) & =  
    \int_{\mathbb{R}^2} dxdy
    \mathcal{A}^{l}_{\mathbf{k}}(x) \mathcal{A}^{l}_{\mathbf{k}}(y)
    % \nonumber \\ &
    \frac{ x - y }
    {x-y + \hbar\omega + i0^{+}} .
    \label{eq:Ik}
  \end{align}
  The damped BECs can be written as
  \begin{align}
    % &\scalebox{.85}{$\displaystyle 
    &\bar{Z}^{\text{}}_{s,\alpha\beta}(\omega,\Sigma^{\text{el}}) 
    \nonumber\\    &
    = \frac{2}{N_\mathbf{k}}
    \sum^{N_\mathbf{k}}_{\mathbf{k},l}
    f'(\varepsilon^{l}_{\mathbf{k}})
    M^{1,s\alpha\beta}_{{\mathbf {k}}ll}
    I^{l}_{\mathbf{k}}(\omega,\Sigma^{\text{el}})
    +  \bar{Z}^{\text{Dyn}}_{s,\alpha\beta}(\omega)
    % $}
    .
  \end{align}

  Considering that the integral of equation (\ref{eq:Ik})  is mostly determined in the region $|x|\sim\Gamma_{\mathbf{k}}$ and $|x-y| \sim2\Gamma_{\mathbf{k}}$, we find that the limits $\frac{\omega}{2\Gamma_{\mathbf{k}}}\rightarrow \infty$ (undamped) and $\frac{\omega}{2\Gamma_{\mathbf{k}}}\rightarrow 0$ (overdamped) coincide respectively to $\bar{Z}^{\text{Dyn}}_{s,\alpha\beta}(\omega)$ and $\bar{Z}^{\text{Stat}}_{s,\alpha\beta}$, since
  \begin{align}
    \lim_{\hbar\omega/2\Gamma_{\mathbf{k}} \rightarrow \infty} 
    I^{l}_{\mathbf{k}}(\omega,\Sigma^{\text{el}}) &= 0 \qquad \text{undamped} \leftrightarrow \text{Dynamic},
    \\
    \lim_{\hbar\omega/2\Gamma_{\mathbf{k}} \rightarrow 0} 
    I^{l}_{\mathbf{k}}(\omega,\Sigma^{\text{el}}) &=1 \qquad \text{overdamped} \leftrightarrow \text{Static},
    \label{eq:static}
  \end{align}    
  where the limit of equation (\ref{eq:static}) is valid only in the normal phase. We further notice that the static limit does not dependent on the amount of scattering, as expected from a macroscopic response  \cite{pines_theory_1989} and that a finite lifetime smooths out the discontinuity in the phase-space origin 
   (Supplementary Section \ref{si:limits}).

  The adimensional function $I^{l}_{\mathbf{k}}(\omega,\Sigma^{\text{el}})$ acquires a simple form 
  %Finally, 
  if the energy dependence of the self-energy is negligible. Indeed,  if $\Sigma^{\text{el}}_{\mathbf{k}l}(x+\varepsilon^{l}_{\mathbf{k}}+i0^+)\simeq\Sigma^{\text{el}}_{\mathbf{k}l}(\varepsilon^{l}_{\mathbf{k}}+i0^+)$, 
  then 
  $\mathcal{A}^{l}_{\mathbf{k}}(x) 
  = \frac{1}{\pi} 
  \frac{\Gamma^{l}_{\mathbf{k}}}{ (x-\Delta^{l}_{\mathbf{k}})^2 + (\Gamma^{l}_{\mathbf{k}})^2 }$, with $\Gamma^{l}_{\mathbf{k}}=-\mathrm{ Im}[\Sigma^{\text{el}}_{\mathbf{k}l}(\varepsilon^{l}_{\mathbf{k}}+i0^+)]$, $\Delta^{l}_{\mathbf{k}}=\rm {Re}[\Sigma^{\text{el}}_{\mathbf{k}l}(\varepsilon^{l}_{\mathbf{k}})]$ and equation~(\ref{eq:Ik}) becomes:
  \begin{align}
    I^{l}_{\mathbf{k}}(\omega,\Sigma^{\text{el}})= 
    \frac{i2\Gamma^{l}_{\mathbf{k}}} { \hbar\omega + i2\Gamma^{l}_{\mathbf{k}}  }.
    \label{eq:dressing_I_K}
  \end{align}

    \vspace{2ex}\noindent  \textbf{Damped BECs within the isotropic Extended-Drude model} \newline
  \noindent  The relation between the BECs and the isotropic Extended-Drude model equation (\ref{eq:BECwithDrude0},\ref{eq:BECwithDrude1}) is obtained generalizing the extended-Drude recipe to a linear response independently on the particular matrix elements. 
  The key step is the factorization of the sum of the product of the vertices and the bubble terms (as in equation (\ref{eq:rho_factorized})) in two separated sums. 
  This is attained assuming that $B^0_{\mathbf{k},ll}$ is weakly anisotropic in $\bf k$, so that it can be approximated by its momentum average. In addition, since $B^0_{\mathbf{k},ll}$ is non-zero only around the Fermi level, the vertices  contribution can be averaged over the Fermi surface, namely:
  \begin{align}
    &\bar{Z}^{\text{}}_{s,\alpha\beta}(\omega,\Sigma^{\text{el}}) -\bar{Z}^{\text{Dyn}}_{s,\alpha\beta}(\omega) =    \frac{2}{N_\mathbf{k}}\sum_{\mathbf{k},l}^{N_\mathbf{k}}
    M^{1,s\alpha\beta}_{\mathbf{k},ll}  
    B^0_{\mathbf{k},ll}(\omega,\Sigma^{\text{el}}) \nonumber \\
    &\simeq  
    \Bigg(
    \frac{2}{N_\mathbf{k}}\sum_{\mathbf{k},l}^{N_\mathbf{k}} 
    \frac{-f'(\varepsilon^{l}_{\mathbf{k}})}{n_{\varepsilon_F}}M^{1,s\alpha\beta}_{\mathbf{k},ll}\Bigg)
    \;
    \Bigg(
    \frac{2}{N_\mathbf{k}}\sum_{\mathbf{k},l}^{N_\mathbf{k}} 
    B^0_{\mathbf{k},ll}(\omega,\Sigma^{\text{el}}) \Bigg)
    \nonumber \\
    &= 
    \Delta\bar{Z}_{s,\alpha\beta} {I}(\omega,\Sigma^{\text{el}}),
    \label{eq:extDrude_factors}
  \end{align}
  where the adimensional isotropic factor ${I}(\omega,\Sigma^{\text{el}})$ is
  \begin{align}
    {I}(\omega,\Sigma^{\text{el}})&=
    -\frac{2}{n_{\varepsilon_F} N_\mathbf{k}}
    \sum_{\mathbf{k},l}^{N_\mathbf{k}} B^0_{\mathbf{k},ll}(\omega,\Sigma^{\text{el}})
    .
    \label{eq:dress_factor_def}
  \end{align}
  The BECs can then be written as
  \begin{align}
    \bar{Z}^{\text{}}_{s,\alpha\beta}&(\omega, \Sigma^{\text{el}}) =  \bar{Z}^{\text{Dyn}}_{s,\alpha\beta} (\omega) 
    +\Delta\bar{Z}_{s,\alpha\beta} 
    {I}(\omega,\Sigma^{\text{el}}) .
    \label{eq:ext_drude_bec}
  \end{align}
  In order to strengthen the analogy with the renown result of the extended-Drude model, we recall that
  \begin{align}
    -i\omega \sigma^{\alpha\beta}_{\text{Drude}}(\omega, \Sigma^{\text{el}}) &= 
    \frac{(\omega^2_p)_{\alpha\beta} }{ 4\pi}
    - \frac{2}{ 4\pi N_\mathbf{k}}
    \sum^{N_\mathbf{k}}_{\mathbf{k},l} 
    J^{\alpha\beta}_{\mathbf{k},ll} B^0_{\mathbf{k},ll}(\omega,\Sigma^{\text{el}}) ,
  \end{align}
  where the matrix elements are
  $
  J^{\alpha\beta}_{\mathbf{k},lm} = \frac{e^2}{\Omega}
\big\langle u^{l}_{\mathbf k}\big| v^{\alpha}_{\mathbf k }\big| u^{m}_{\mathbf k }\big\rangle 
  \big\langle u^{m}_{\mathbf k }\big| v^{\beta }_{\mathbf k }\big| u^{l}_{\mathbf k}\big\rangle
  $ and the diamagnetic term is expressed by the  plasma frequency.
If the same steps used for obtaining equation (\ref{eq:ext_drude_bec}) are followed for the conductivity, we find the well-known result  \cite{nam_theory_1967}\cite{shulga_electronic_1991}:
  \begin{align}
    -i\omega \sigma^{\alpha\beta}_{\text{Drude}}(\omega, \Sigma^{\text{el}}) &=  
    \frac{(\omega^2_p)_{\alpha\beta}}{ 4\pi} \left[1 - {I}(\omega,\Sigma^{\text{el}}) \right].
    \label{eq:sigmadrude}
  \end{align}
  Combining this with equation \eqref{eq:ext_drude_bec} we finally obtain equations \eqref{eq:BECwithDrude0},\eqref{eq:BECwithDrude1}.
  In the SC phase we assumed that both $\Delta\bar{Z}_{s,\alpha\beta}$ and the optical conductivity can be expressed in terms of the same adimensional isotropic factor ${I}(\omega,\Sigma^{\text{el}})$ as given by equation (\ref{eq:BECwithDrude1}), with $\sigma_{\text{Drude}}(\omega,\Sigma^{\text{el}})$ evaluated in the superconducting state as detailed in Supplementary Section VI.

  \section*{DATA AVAILABILITY}\label{competing}
The Eliashberg spectral function and phonon frequencies, displacements
and lifetimes including anharmonic effects as published in Ref. \cite{bianco_high-pressure_2018}
have been provided by R. Bianco. Experimental data from Ref.  \cite{capitani_spectroscopic_2017} have
been provided by E.J. Nicol. All other data are available from the
corresponding authors upon reasonable request.

  \section*{CODE AVAILABILITY}\label{competing}
The QUANTUM ESPRESSO code used in this research is open source and
available at www.quantum-espresso.org. We used an in-house customized
version of QUANTUM ESPRESSO with the additional computational
capabilities discussed in this work, that is available from G.Marchese
upon reasonable request. Wannier interpolation and solution of
Migdal-Eliashberg equations have been performed using the open-source
software EPIq \cite{marini_epiq_2023}.

%\clearpage

%\textbf{Methods-only references}
%\bibliographyMethods{h3s_prl}

  \onecolumngrid

  \clearpage
  \section*{Extended data}
  \setcounter{figure}{0} 
  \setcounter{table}{0} 
  \renewcommand{\figurename}{}
  \renewcommand{\thefigure}{\textbf{Extended Data Fig. \arabic{figure}}}
  \renewcommand{\tablename}{}
  \renewcommand{\thetable}{\textbf{Extended Data Table \arabic{table}}}

  \begin{table}[h]
    \begin{tabular}{cc p{1ex}||p{1ex} c|c|c|c|c p{1ex}||p{1ex} c|c|c}
      \multicolumn{4}{c}{System}&\multicolumn{5}{c}{Static}&\multicolumn{2}{c}{}& \multicolumn{3}{c}{Dynamic}\\
      Formula                 & P (GPa) &&& $V$ & $T$ & $V+T$ &  $\sum_s \bar{Z}^{\text{Stat}}_{ s,\alpha \beta}$  & \% err. &&& $\frac{m\Omega}{4 \pi e^2 }  (\omega_{\text{p}}^2)^{Z}_{\alpha\beta} $  &   $\sum_s \bar{Z}^{\text{Dyn}}_{ s,\alpha \beta}(i0^+)$  & \% err. \\
      \midrule
      H$_3$S  & 130  &&& $0.007$ & $9.858$ & $9.865$ & $9.867$   & $0.028$ &&& $1.972$ & $2.015$ & $2.134$ \\
      H$_3$S  & 150  &&& $0.141$ & $10.056$ & $10.197$ & $10.199$   & $0.017$ &&& $1.983$ & $2.028$ & $2.249$ \\
      H$_3$S  & 200  &&& $0.437$ & $10.487$ & $10.924$ & $10.926$   & $0.019$ &&& $2.000$ & $2.048$ & $2.380$\\[1ex]
    Al  & 0  &&& $3.288$ & $3.196$ & $6.484$ & $6.482$ & $0.030$ &&& $1.986$ & $1.987$ & $0.017$ \\
    Al  & 100  &&& $1.975$ & $2.742$ & $4.717$ & $4.715$ & $0.040$ &&& $1.555$ & $1.555$ & $0.024$ \\
    \end{tabular}
    \caption{\textbf{Numerical validation of the static BECs sum rule.}
    Numerical values of the terms appearing in equation (\ref{eq_main:adiab_sumrule}) are reported for H$_3$S and Al as computed by the DFPT calculation described in Supplementary Section \ref{si:calc_param}.
    We introduce the following shorthand notation in the table: $V$ stands for $-n_{\varepsilon_{F}}  \sum_{ s} \langle{\frac{\partial \bar{V}^{\mathbf{\Gamma} }_{s\beta }}{i\partial q_{\alpha}}}\rangle_F$ and $T$ for $n_{\varepsilon_{F}} \langle{ v^{\alpha}_{\mathbf{k}} p_{\mathbf k}^{\beta} }\rangle_F$. As a comparison, we provide numerical results for the dynamical BECs sumrule in the dynamic limit,
    %the test for the dynamic limit sumrule 
    equation (\ref{eq:dynsumrule}), at equal convergence parameters.
    The values of the individual atomic contributions to the sum rule of H$_3$S are listed in the \ref{tab:h3s_sumrule_comps}. The largest errors are observed in the dynamic sum rule for H$_3$S and are due to the finite imaginary constant of the $i0^+$ regularization in $\bar{Z}^{\text{Dyn}}_{ s,\alpha \beta}(i0^+)$ (see Sec. \ref{si:calc_param}). A finite  $i0^+$  constant should be used only in the dynamical case and in presence of low-energy vertical interband transitions, that occur in the H$_3$S band-structure, but not in the Al one.}
    \label{tab:h3s_al_sumrule}
  \end{table}

  \begin{table}[h]
    \begin{center}
      \begin{tabular}{@{}c|c|c|c@{}}
        \toprule
          & $Z^S $ & $ Z_{\parallel}^H $ & $ Z_{\perp}^H $ \\
      \midrule
$\bar{Z}^{\text{Dyn}} (i0^+)$ & $-1.306$ & $3.744$ & $-0.205$\\

$ \bar{Z}_{s,\alpha\beta}^{\text{Stat}}$ & $8.423$ & $0.007$ & $0.884$\\

$ \Delta\bar{Z}_{s,\alpha\beta}$ & $9.729$ & $-3.737$ & $1.089$\\
  \midrule
$\bar{Z}_{s,\alpha\beta}^{\text{Dyn}}(\omega_{1} )$ & $-1.313-i0.055$ & $3.749+i0.039$ & $-0.204+i0.001$\\

$\bar{Z}_{s,\alpha\beta}^{\text{Dyn}}(\omega_{2} )$ & $-1.326-i0.099$ & $3.760+i0.071$ & $-0.204+i0.002$\\
  \midrule
$\bar{Z}_{s,\alpha\beta}^{\text{}}(\omega_{1},\Sigma^{\text{el}},300  \text{ K} )$ & $6.775+i1.502$ & $0.643-i0.559$ & $0.701+i0.175$\\

$\bar{Z}_{s,\alpha\beta}^{\text{}}(\omega_{2},\Sigma^{\text{el}},300  \text{ K} )$ & $6.337+i1.466$ & $0.817-i0.530$ & $0.654+i0.177$\\
  \midrule
$\bar{Z}_{s,\alpha\beta}^{\text{}}(\omega_{1},\Sigma^{\text{el}},20  \text{ K} )$ & $6.680-i0.044$ & $0.679+i0.035$ & $0.690+i0.002$\\

$\bar{Z}_{s,\alpha\beta}^{\text{}}(\omega_{2},\Sigma^{\text{el}},20  \text{ K} )$ & $6.166+i1.011$ & $0.882-i0.355$ & $0.635+i0.126$\\
        \botrule
      \end{tabular}
      \caption{\textbf{Comparison of Born effective charges.}
      This table collects the values of BECs obtained as described in Supplementary Section \ref{si:becs} for the $Im\bar{3}m$ phase of H$_3$S at 150 GPa. 
      The cubic symmetry of the crystal implies that only three components of the BEC tensors are independent: one for the sulfur atom ($Z^S$) and two for the $H$-atom, corresponding respectively to the ion displacement along the H-S bond ($Z^H_{\parallel}$) or perpendicular to it ($Z^H_{\perp}$). We list BECs in the dynamic and static limits as well as their values at the phonon frequencies of $\omega_1=84$  and $\omega_2=148$  meV,  comparing undamped $\bar{Z}_{s,\alpha\beta}^{\text{Dyn}}(\omega_{\mu} )$ with damped $\bar{Z}_{s,\alpha\beta}^{\text{}}(\omega_\mu,\Sigma^{\text{el}})$ BECs that include self-energy effects via equation  %the inclusion of self-energy effects via equation 
      (\ref{eq:BECwithDrude0}). Damped BECs are evaluated both in the normal ($T=300$~K) and SC ($T=20$~K) state. Notice that $\Im\bar{Z}_{s,\alpha\beta}(\omega_1,\Sigma^{\text{el}},20K)\sim 0$.  
      }
      \label{tab:h3s_dirty_bec_150-cubic}%
    \end{center}
  \end{table}

  \begin{figure}[h]%
    \centering
    \includegraphics[width=160mm]{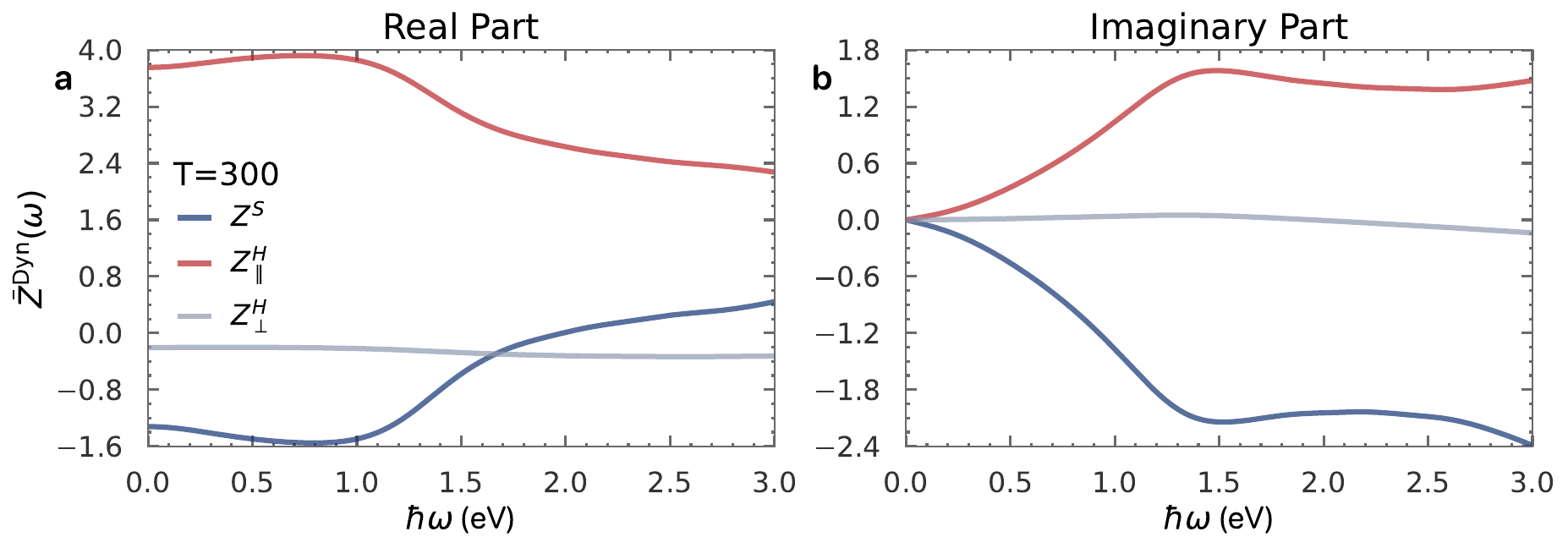}
    \caption{
      \textbf{Undamped BECs in a wider frequency range.} The independent components of the undamped dynamical BECs are plotted on their typical variation scale, which is larger than the vibrational frequency range of \ref{fig:chargefig}. In particular, we can notice a strong variation for $\hbar \omega > 1$eV due to the onset of interband transitions. This onset is coherent with the reflectivity drop observed at the same energy (figure \ref{fig:R_excitations}\textbf{a}).
    }
    \label{fig:dyn_becs}
  \end{figure}

  \begin{figure}[h]%
    \centering
    \includegraphics[width=80mm]{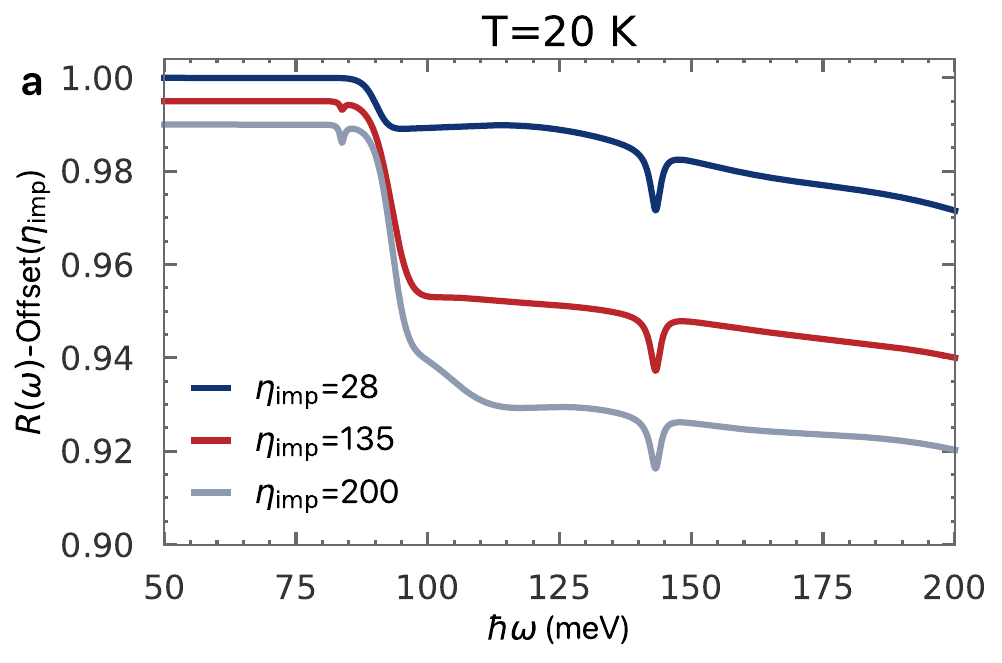}~
    \includegraphics[width=80mm]{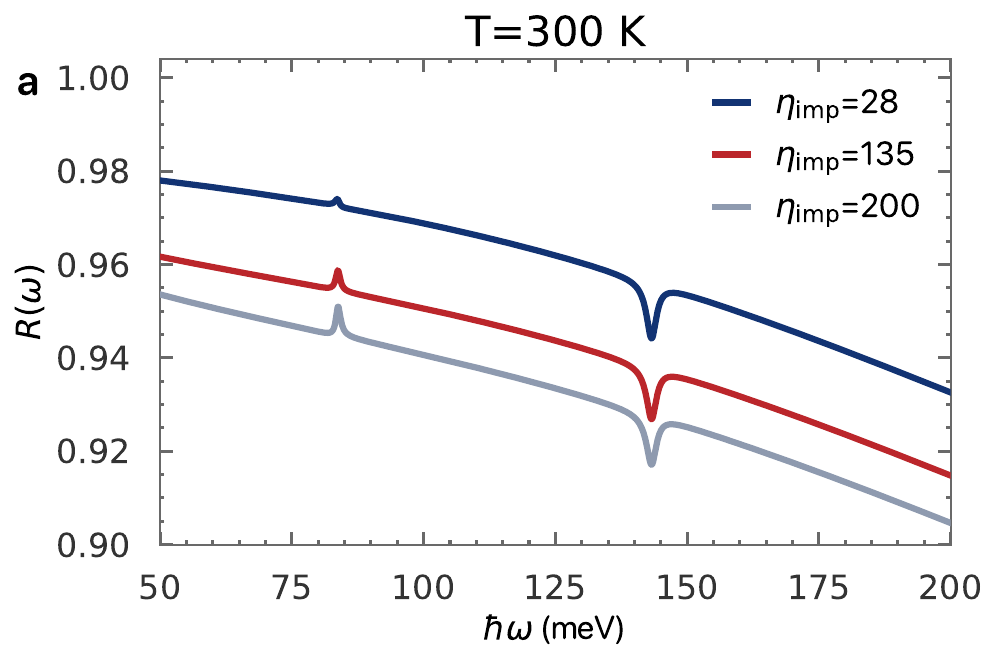}~
    \caption{ \textbf{Reflectivity spectra at different impurity scattering rates} (\textbf{a}) Superconducting and (\textbf{b}) normal phase reflectivity spectra, where the self-energy is evaluated with the procedure explained in the Supplementary Section \ref{si:extdrude} for three different impurity scattering rates $\eta_{\text{imp}}$. In the left panel the offset is 0 for $\eta_{\text{imp}}$=28 meV, 0.01 for $\eta_{\text{imp}}$=135 meV and 0.02 for $\eta_{\text{imp}}$=200 meV. The impurity scattering rate affects differently the electronic baseline and the vibrational resonances. In the superconducting phase higher impurity scattering rates correspond to sharper drops of the reflectivity for $\omega\sim 2\Delta$, while in the normal phase they correspond to vertical shifts of the electronic baseline. The intensity of the vibrational resonances is enhanced by an increase of impurity scattering rates in both cases, while their shape is affected in a less trivial way due to the interplay between electronic and lattice responses, as manifested by the different effects upon the two resonances centered around 84 and 148 meV.
    }
    \label{fig:R_vs_imp}
  \end{figure}

\clearpage
  \section*{SUPPLEMENTARY INFORMATION}

  \setcounter{figure}{0} 
  \setcounter{table}{0} 
  \setcounter{equation}{0} 
  \renewcommand{\figurename}{}
  \renewcommand{\thefigure}{\textbf{Supplementary Figure \arabic{figure}}}
  \renewcommand{\tablename}{\textbf{}}
  \renewcommand{\thetable}{\textbf{Supplementary Table \arabic{table}}}
  \renewcommand{\theequation}{\textbf{S}\arabic{equation}}

 \renewcommand{\thesubsection}{\Roman{subsection}} 
 
\textbf{\underline{Table of contents}}
\begin{enumerate}
    \item[] \textbf{Additional theoretical  details:}
    \item[\ref{si:vel_op}] Velocity operator, plasma frequency and sum rule for dynamical BECs
    \item[\ref{si:becs_sumrule}.] Proof of Static BECs sum rule
    \item[\ref{si:limits}.] Comparison of static and dynamic limit
    \item[]\textbf{ Computational  details:}
    \item[\ref{si:ingre}.] Ingredients for \textit{ab-initio} reflectivity spectra
    \item[\ref{si:calc_param}.] Computational parameters
    \item[\ref{si:extdrude}.] Electronic response and Extended-Drude calculation
    \item[\ref{si:vibr}.] Lattice vibrations
    \item[\ref{si:becs}.] Numerical calculation of Born Effective Charges  
\end{enumerate} 

 \subsection{\label{si:vel_op}Velocity operator, plasma frequency and sum rule for dynamical BECs}
  The use of non-local pseudopotentials in numerical calculations introduces a non-local contribution to the DFT Hamiltonian. As a consequence, the use of the velocity operator requires some additional care. This operator is defined as
  \begin{align}
    \bm{v}_{\mathbf{k}} =  \frac{\partial H_{\mathbf{k}}}{\hbar\partial \mathbf{k}}
    = \frac{\mathbf{p}_{\mathbf{k}}}{m}  +   \frac{\partial V_{\mathbf{k}}^{\text{nl}}}{\hbar\partial{\mathbf{k}} },
  \end{align}
  where $ V_{\mathbf{k}}^{\text{nl}}$ stands for the non-local part of the pseudopotential.  The second term in the r.h.s. of the previous equation is not present in the all-electron case where we can identify $\bm{v}_{\mathbf{k}}$ with ${\mathbf{p}_{\mathbf{k}}}/{m}$. In the derivations and the expressions of effective charges reported in this paper we keep the distinction between $\bm{v}_{\mathbf{k}}$ and ${\mathbf{p}_{\mathbf{k}}}/{m}$. 
    In particular, two expressions for the plasma frequency are employed. The first:
   \begin{align}
    (\omega_{\text{p}}^2)_{\alpha\beta}^{Z} &= 
    \frac{  4 \pi e^2  }{\Omega }  \frac{2}{N_\mathbf{k}}\sum^{N_\mathbf{k}}_{\mathbf{k},l} 
    -f'(\varepsilon^{l}_{\mathbf{k}}) 
    \big\langle u^{l}_{\mathbf k}\big| v^{\alpha}_{\mathbf{k}}\big| u^{l}_{\mathbf k}\big\rangle 
    \bigg\langle{u^{l}_{\mathbf k}}\bigg|{\frac{ p^{\beta}_{\mathbf{k}} }{m}}\bigg|{u^{l}_{\mathbf k}}\bigg\rangle
    \label{eq:plasma4becs}
  \end{align}
  should be used 
  for the dynamical BECs sum rule proven in Refs \cite{dreyer_nonadiabatic_2022,resta_theory_2022}, that is :
  \begin{align}
\sum_s  \bar{Z}^{\text{Dyn}}_{s,\alpha\beta}(i0^+) &=   \frac{m\Omega  }{4 \pi e^2}   (\omega_{\text{p}}^2)^{Z}_{\alpha\beta}.
    \label{eq:dynsumrule}
  \end{align}
On the other hand, only the velocity operator should be used for the calculation of the plasma frequency entering in the electronic optical conductivity:
  \begin{align}
    (\omega_{\text{p}}^2)_{\alpha\beta}^{\sigma} &=  
    \frac{  4 \pi e^2 }{\Omega }  \frac{2}{N_\mathbf{k}}\sum^{N_\mathbf{k}}_{\mathbf{k},l} 
    -f'(\varepsilon^{l}_{\mathbf{k}}) 
    \big\langle u^{l}_{\mathbf k}\big| v^{\alpha}_{\mathbf{k}}\big| u^{l}_{\mathbf k}\big\rangle
    \big\langle u^{l}_{\mathbf k}\big| v^{\beta}_{\mathbf{k}}\big| u^{l}_{\mathbf k}\big\rangle.
    \label{eq:plasma4cond}
  \end{align}
  Indeed, we use this expression for the calculation of $\chi^{\rm el}$ in equation~(\ref{eq:epsilontot}).
 %In  \ref{tab:h3s_al_vel_vs_p} we  compare these two definitions for the plasma frequencies numerically.
Numerical results of these two definitions for the plasma frequency are compared in  \ref{tab:h3s_al_vel_vs_p}.

  \begin{table}[h]
    \begin{tabular}{ccc p{1ex}||p{1ex} c|c p{1ex}||p{1ex} c|c}
      Formula                 & P (GPa) &  $\Omega$ ($a_0^3$)
      &&&  $\frac{m\Omega}{4 \pi e^2 } (\omega_{\text{p}}^2)_{\alpha\alpha}^{\sigma}  $  & $\frac{m\Omega}{4 \pi e^2 }  (\omega_{\text{p}}^2)_{\alpha\alpha}^{Z} $ 
      &&&  $ \hbar \omega_{\text{p}}^{\sigma}$ (eV)   & $\hbar \omega_{\text{p}}^{Z}$ (eV)  \\
      \midrule
      H$_3$S    & 130 & 100.8 &&&  1.83  & 1.97  &&& 13.01  &13.49  \\
      H$_3$S    & 150 &  97.1 &&&  1.84  & 1.98  &&& 13.29  &13.79 \\
      H$_3$S    & 200 &  89.7 &&&  1.86  & 2.01  &&& 13.86  &14.40 \\[1ex]
      Al        & 0   & 112.6 &&&  1.87  & 1.99  &&& 12.44  & 12.81  \\
      Al        & 100 & 69.5 &&&  1.40  & 1.55  &&& 13.68  & 14.43 \\
    \end{tabular}
    \caption{\textbf{Non-local contribution and plasma frequency.} Comparison of the two expressions to be used for the calculation of the plasma frequency in presence of non-local pseudo-potentials. We report the values for different compounds (H$_3$S and Al) at different pressures (130,150,200 and 0,100 GPa) and unit cell volume $\Omega$. The plasma frequencies appear both in eV and in the adimensional unit for the sumrule evaluation.  }
    \label{tab:h3s_al_vel_vs_p}
  \end{table}

  \subsection{\label{si:becs_sumrule}Proof of Static BECs sum rule}

In this section we find a simple expression for the sum of the static BECs over the ions. We can take advantage of the dynamical BECs sum rule  (equation  (\ref{eq:dynsumrule})):
  \begin{align}
    \sum_s  \bar{Z}^{\text{Stat}}_{s,\alpha\beta} &= 
    \sum_s  \bar{Z}^{\text{Dyn}}_{s,\alpha\beta}  +  \sum_s \Delta \bar{Z}_{s,\alpha\beta}
    = \frac{m \Omega}{4 \pi e^2}   (\omega_{\text{p}}^2)^{Z}_{\alpha\beta}  +  \sum_s \Delta \bar{Z}_{s,\alpha\beta},
  \end{align}
  and focus only on the sum over the atoms of the discontinuity step defined in equation (\ref{eq_main:singul_step}):
  \begin{align}
    &\Delta \bar{Z}_{s,\alpha\beta} =
    % \nonumber\\&
    - \frac{2}{N_\mathbf{k}}\sum^{N_\mathbf{k}}_{\substack{\mathbf {k},l,m\\m\ne l}}   f^{'}(\varepsilon_{\mathbf{k}}^{l})   
    \bigg\langle{u^{l}_{\mathbf k}}\bigg|{ \frac{i\hbar v^{\alpha}_{\mathbf{k}}}
    {\varepsilon_{\mathbf{k}}^{l} -\varepsilon_{\mathbf{k}}^{m} +i0^{+}}  }\bigg|{u^{m}_{\mathbf k} }\bigg\rangle 
    \big\langle u^{m}_{\mathbf k}\big|  \bar{V}^{\mathbf{\Gamma} }_{s\beta } \big| u^{l}_{\mathbf k} \big\rangle
    % \nonumber \\&
    +  \frac{2}{N_\mathbf{k}}\sum^{N_\mathbf{k}}_{\mathbf{k},l} f^{'}(\varepsilon_{\mathbf{k}}^{l})
    \bigg\langle{u^{l}_{\mathbf k}}
    \bigg|{ \frac{\partial \bar{V}^{\mathbf{\Gamma} }_{s\beta }}{i\partial q_{\alpha}}   }\bigg|
    {u^{l}_{\mathbf k} } \bigg\rangle.
  \end{align}
  The sum of the first term in equation (\ref{eq_main:singul_step})  can be simplified exploiting equation (\ref{eq:g2p}):
  \begin{align}
    - &\frac{2}{N_\mathbf{k}} \sum_s \sum^{N_\mathbf{k}}_{\substack{{\mathbf {k}},l,m\\m\ne l}}   
    f^{'}(\varepsilon_{\mathbf{k}}^{l})   
    \bigg\langle{u^{l}_{\mathbf k}}\bigg|{\frac{i\hbar v^{\alpha}_{\mathbf{k}}}{(\varepsilon_{\mathbf{k}}^{l} -\varepsilon_{\mathbf{k}}^{m})}}\bigg|{u^{l}_{\mathbf k}}\bigg\rangle 
    \big\langle u^{m}_{\mathbf k}\big| \bar{V}^{\mathbf{\Gamma} }_{s,\beta }\big| u^{l}_{\mathbf k}\big\rangle =
    - \frac{2}{N_\mathbf{k}}\sum^{N_\mathbf{k}}_{\substack{{\mathbf {k}},l,m\\m\ne l}}  
    f^{'}(\varepsilon_{\mathbf{k}}^{l})    
    \big\langle u^{l}_{\mathbf k}\big| v^{\alpha}_{\mathbf{k}}\big| u^{m}_{\mathbf k}\big\rangle
    \big\langle u^{m}_{\mathbf k}\big| p^{\beta}_{\mathbf{k}}\big| u^{l}_{\mathbf k}\big\rangle.
    % \nonumber\\
  \end{align}
  We can then rewrite the sum over bands in the following way:
  \begin{align}
    - \frac{2}{N_\mathbf{k}}\sum^{N_\mathbf{k}}_{\substack{{\mathbf {k}},l,m\\m\ne l}}  
    &f^{'}(\varepsilon_{\mathbf{k}}^{l})    
    \big\langle u^{l}_{\mathbf k}\big| v^{\alpha}_{\mathbf{k}}\big| u^{m}_{\mathbf k}\big\rangle
    \big\langle u^{m}_{\mathbf k}\big| p^{\beta}_{\mathbf{k}}\big| u^{l}_{\mathbf k}\big\rangle = \nonumber\\
    &=  - \frac{2}{N_\mathbf{k}}\sum^{N_\mathbf{k}}_{\substack{{\mathbf {k}},l}} 
    f^{'}(\varepsilon_{\mathbf{k}}^{l})   
    \Big\langle{u^{l}_{\mathbf k}}\Big|{v^{\alpha}_{\mathbf{k}}
    \sum_{m}}\Big|{u^{m}_{\mathbf k}}\Big\rangle
    \big\langle u^{m}_{\mathbf k}\big| p^{\beta}_{\mathbf{k}}\big| u^{l}_{\mathbf k}\big\rangle   %\nonumber\\&
    +  \frac{2}{N_\mathbf{k}}\sum^{N_\mathbf{k}}_{\substack{{\mathbf {k}},l}} 
    f^{'}(\varepsilon_{\mathbf{k}}^{l})  
    \big\langle u^{l}_{\mathbf k}\big| v^{\alpha}_{\mathbf{k}}\big| u^{l}_{\mathbf k}\big\rangle
    \big\langle u^{l}_{\mathbf k}\big| p^{\beta}_{\mathbf{k}}\big| u^{l}_{\mathbf k}\big\rangle  \nonumber\\
    &= - \frac{2}{N_\mathbf{k}}\sum^{N_\mathbf{k}}_{\substack{{\mathbf {k}},l}} 
    f^{'}(\varepsilon_{\mathbf{k}}^{l})   
    \big\langle u^{l}_{\mathbf k}\big|  v^{\alpha}_{\mathbf{k}} p^{\beta}_{\mathbf{k}} \big| u^{l}_{\mathbf k} \big\rangle
    -   \frac{m \Omega}{4 \pi e^2} (\omega_{\text{p}}^2)^{Z}_{\alpha\beta}
    =  n_{\varepsilon_F}\langle    v^{\alpha}_{\mathbf{k}} p^{\beta}_{\mathbf{k}}   \rangle_F
    - \frac{m \Omega}{ 4 \pi e^2} (\omega_{\text{p}}^2)^{Z}_{\alpha\beta} ,
  \end{align}
  from which it follows:
  \begin{align}
    \sum_s  \bar{Z}^{\text{Stat}}_{s,\alpha\beta} &= 
    \frac{m \Omega}{4 \pi e^2}   (\omega_{\text{p}}^2)^{Z}_{\alpha\beta}  + \Bigg[n_{\varepsilon_F} \langle    v^{\alpha}_{\mathbf{k}} p^{\beta}_{\mathbf{k}}   \rangle_F
    - \frac{m \Omega}{ 4 \pi e^2} (\omega_{\text{p}}^2)^{Z}_{\alpha\beta} \Bigg] + \sum_s\frac{2}{N_\mathbf{k}}\sum^{N_\mathbf{k}}_{\mathbf{k},l} f^{'}(\varepsilon_{\mathbf{k}}^{l})
    \bigg\langle{u^{l}_{\mathbf k}}\bigg|
    { \frac{\partial \bar{V}^{\mathbf{\Gamma} }_{s\beta }}{i\partial q_{\alpha}}   }
    \bigg|{u^{l}_{\mathbf k} }\bigg\rangle \nonumber\\
    &= 
    n_{\varepsilon_F}\langle    v^{\alpha}_{\mathbf{k}} p^{\beta}_{\mathbf{k}}   \rangle_F
    - n_{\varepsilon_F}\sum_s \langle  \frac{\partial \bar{V}^{\mathbf{\Gamma} }_{s\beta }}{i\partial q_{\alpha}}\rangle_F  .
  \end{align}

  \newpage
  \subsection{Comparison of static and dynamic limit \label{si:limits}}
\begin{figure}[h]%
    \centering
    \includegraphics[width=70mm]{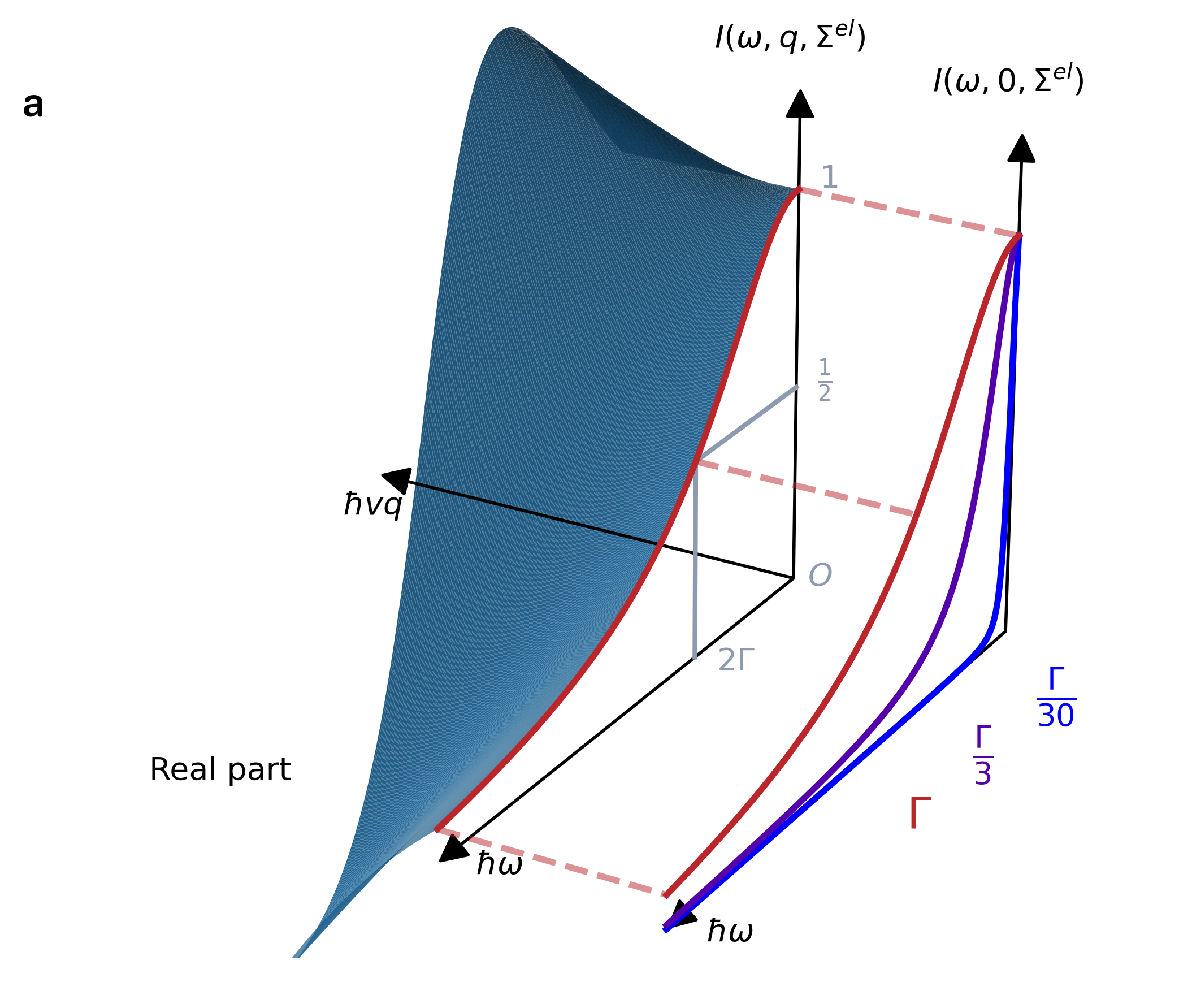}~
    \includegraphics[width=70mm]{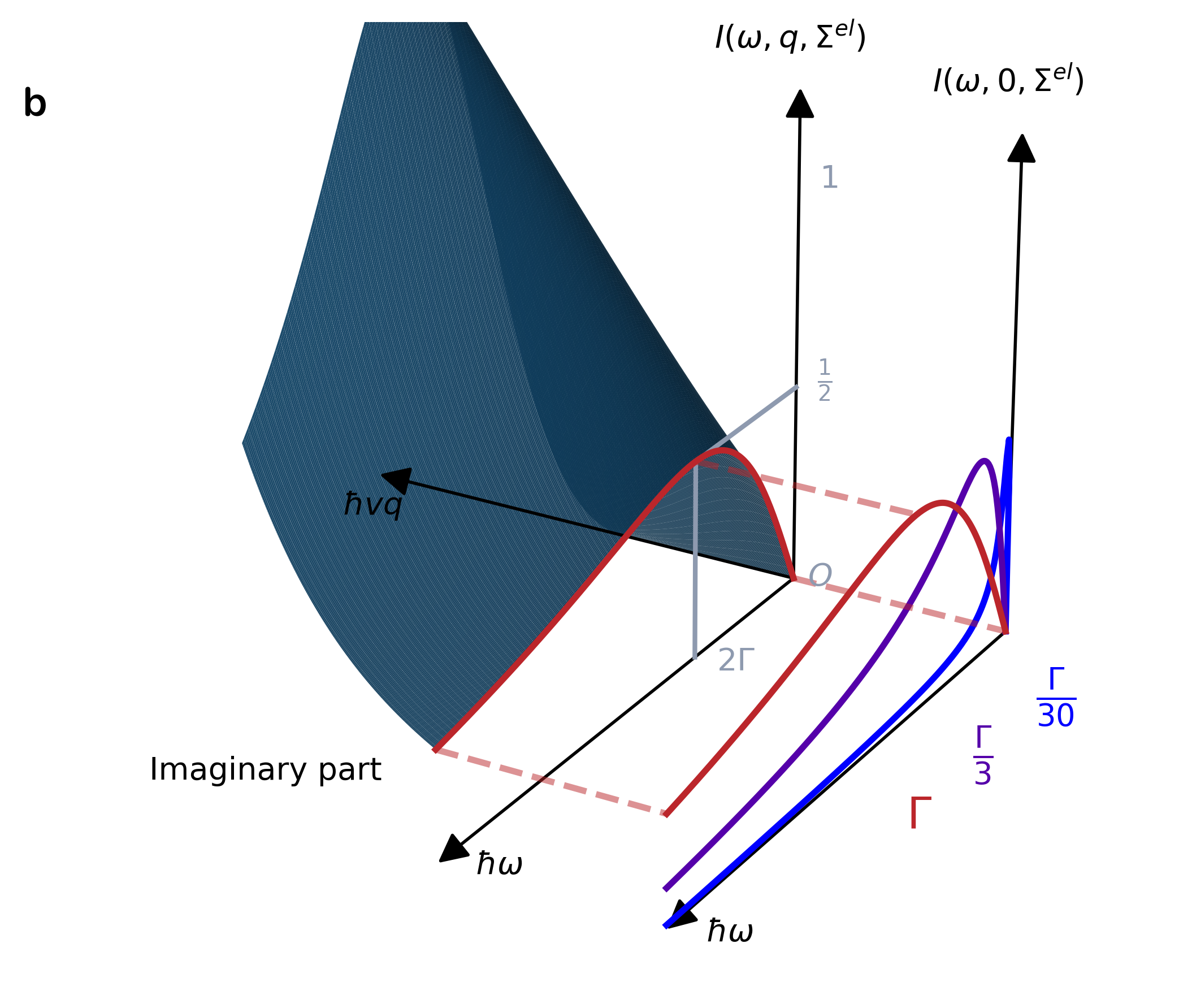}~
    \caption{
      \textbf{Electron-hole propagator singularity smoothed out by scattering}.
      The adimensional factor $I_{\mathbf{k}}(\omega,\mathbf{q},-i\Gamma_{\mathbf{k}})$ describing the electron-hole propagator within the Lorentzian extended-Drude approach [equation (\ref{eq:extD_finite_q})] is plotted as a function of frequency, momentum and self-energy value. Since the plot is valid for each $\mathbf{k}$, we drop such index, and call $\hbar\bm{v_{k}}\cdot \mathbf{q}=\hbar vq$. It can be clearly seen that a finite scattering rate smooths out the singular behavior of the electron-hole propagator in the phase-space origin. As soon as $\Gamma\rightarrow 0$ goes to zero the discontinuity appears along the dynamic limit $(\omega\rightarrow 0, 0)$ in both the real (\textbf{a}) and imaginary (\textbf{b}) parts.
      This behavior is captured by the plot on the right side of each panel depicting the dynamic limit $I(\omega,0,-i\Gamma)$ for decreasing scattering values ($\Gamma,\frac{\Gamma}{3},\frac{\Gamma}{30}$) .
    }
    \label{fig:el-hole}
  \end{figure}

  In this section we present a simple case that exemplifies the singular behaviour of the electron-hole propagator in the phase-space origin for the normal phase.  Following the same steps presented in the Methods Section, we study the electron-hole propagator at small momentum $\bf q$, so that $\varepsilon^{l}_{\mathbf{k}+\mathbf{q}}\simeq \varepsilon^{l}_{\mathbf{k}} + \hbar  \bm{v}^{l}_{\mathbf{k}}\cdot\mathbf{q}$ and
  \begin{align}
    B_{\mathbf{k},ll}(\omega,\mathbf{q},\Sigma^{\text{el}})
    =
    \int_{\mathbb{R}^2} dx dy
    \mathcal{A}^{l}_{\mathbf{k}}(x) \mathcal{A}^{l}_{\mathbf{k}}(y)
    % \nonumber\\&
    \frac{f(\varepsilon^{l}_{\mathbf{k}}+x)-  f(\varepsilon^{l}_{\mathbf{k}}+\hbar\bm{v}^{l}_{\mathbf{k}}\cdot\mathbf{q}+y)}
    {x -  \hbar\bm{v}^{l}_{\mathbf{k}}\cdot\mathbf{q}-y + \hbar \omega + i0^{+}}.
  \end{align}
In the high-temperature limit, $f(\varepsilon^{l}_{\mathbf{k}}+x)-f(\varepsilon^{l}_{\mathbf{k}}+\hbar\bm{v}^{l}_{\mathbf{k}}\cdot\mathbf{q}+y)\simeq f'(\varepsilon^{l}_{\mathbf{k}})(x-\hbar\bm{v}^{l}_{\mathbf{k}}\cdot\mathbf{q}-y)$  and assuming Lorentzian spectral functions, we can generalize to finite momenta the expression for the adimensional dressing function given in equation (\ref{eq:dressing_I_K}); as in  \cite{maksimov_nonadiabatic_1996}, we obtain  $B_{\mathbf{k},ll}(\omega,\mathbf{q},\Sigma^{\text{el}})\simeq f'(\varepsilon^{l}_{\mathbf{k}}) I_{\mathbf{k}}(\omega,\mathbf{q},-i\Gamma_{\mathbf{k}})$, where
  \begin{align}
    I_{\mathbf{k}}(\omega,\mathbf{q},-i\Gamma_{\mathbf{k}})
    =
    \frac{i2\Gamma_{\mathbf{k}} - \hbar\bm{v}^{l}_{\mathbf{k}}\cdot\mathbf{q}}
    {\hbar \omega  - \hbar\bm{v}^{l}_{\mathbf{k}}\cdot\mathbf{q} + i2\Gamma_{\mathbf{k}}}.
    \label{eq:extD_finite_q}
  \end{align}    
  %We represent the values of t
  The above function is displayed in   \ref{fig:el-hole}. It is straightforward to see that a step-like discontinuity appears when approaching the origin  in the absence of scattering $\Gamma_{\mathbf{k}}=0$ along $(\omega,\mathbf{0})$ or $(0,\mathbf{q})$ namely:
  \begin{align}
    \lim_{\omega\rightarrow0} I_{\mathbf{k}}(\omega,\mathbf{0},0) = 0 \\
    \lim_{q\rightarrow0} I_{\mathbf{k}}(0,\mathbf{q},0) = 1 .
  \end{align} 
  Instead, for any finite $\Gamma_{\mathbf{k}}$, the function  $I_{\mathbf{k}}(\omega,\mathbf{q},-i\Gamma_{\mathbf{k}})$ is continuous in the origin of the $(\omega,\mathbf{q})$ space and:
   \begin{align}
    \lim_{\omega\rightarrow0} I_{\mathbf{k}}(\omega,\mathbf{0},-i\Gamma_{\mathbf{k}}) =
    \lim_{q\rightarrow0} I_{\mathbf{k}}(0,\mathbf{q},-i\Gamma_{\mathbf{k}}) = 1 .
  \end{align}

  \subsection{\label{si:ingre} Ingredients for \textit{ab-initio} reflectivity spectra}

  As discussed in the main text, it is possible to distinguish the electronic and lattice  response to the impinging electromagnetic beam.

  The electronic response in conventional metals is characterized by the adiabatic inertia \cite{resta_drude_2018} quantified with the plasma frequency $\omega_{\text{p}}$,
  by the electron-hole transitions accounted for with the interband optical conductivity $\sigma_{\text{Inter}}(\omega,\Sigma^{\text{el}})$ and by the presence of electronic scattering which can be represented via the adimensional function $I(\omega,\Sigma^{\text{el}})$ within the extended-Drude procedure.
  We obtain the first two quantities $\omega_{\text{p}}$, $\sigma_{\text{Inter}}(\omega,\Sigma^{\text{el}})$ thanks to DFPT calculation (whose parameters are reported in Section \ref{si:calc_param}) and using an interpolation scheme based on Wannier functions \cite{marzari_maximally_1997,souza_maximally_2001} as implemented in the EPIq software \cite{marini_epiq_2023}.
  Section \ref{si:extdrude} provides more details on how we compute the adimensional function $I(\omega,\Sigma^{\text{el}})$ considering electron-phonon and impurity scattering through the solution of  Migdal-Eliashberg equations.
   
The vibrational contribution to the dielectric response requires both the knowledge of lattice vibrations and of the coupling between ion displacements and external field.
Section \ref{si:vibr} specifies how we treat the dynamical matrices describing energy, polarization and lifetime of lattice vibrations.
Section \ref{si:becs} describes how we compute the BECs in the dynamical and static cases.
We include scattering effects on the BECs using the same adimensional function $I(\omega,\Sigma^{\text{el}})$ evaluated for the electronic response, as discussed in the Method Section.

  \subsection{\label{si:calc_param}Computational parameters}  
  \subsubsection{H$_3$S}
  The system at the equilibrium is simulated via a DFT calculation, as implemented in the Quantum Espresso package  \cite{giannozzi_quantum_2009,giannozzi_advanced_2017}, with a plane-wave energy cutoff of 90 Ry and a density cutoff of 360 Ry over a Monkhorst-Pack of $48^3$ points displaced from the Brillouin zone origin. 
  A Gaussian smearing temperature of 408 meV has been adopted. The cell has a BCC $Im\bar{3}m$ structure at all pressures.
  The lattice parameter for the different pressures was evaluated via standard DFT relaxation procedure with a threshold of 0.05 GPa, giving 5.86, 5.79,5.64 Bohr  for pressures of 130, 150, 200 GPa, respectively.

  We compute $ n_{\varepsilon_{F}} \langle v^{\alpha}_{\mathbf{k}} p_{\mathbf k}^{\beta} \rangle_F$ and the plasma frequency using both definitions of equations (\ref{eq:plasma4becs},\ref{eq:plasma4cond})  directly from the DFT calculation.
  To obtain the static BECs, we run a DFPT calculation on top of the previous DFT results as explained in \cite{macheda_2022}.
  For the dynamical BECs and interband conductivity we act differently. The $\mathbf k$-mesh requirements for metallic systems and the number of frequencies needed to realize a convergent and smooth  reflectivity spectrum prevent a  calculation through a straightforward self-consistent time dependent DFPT (TD-DFPT) calculation  \cite{binci_first-principles_2021}. Hence, we adopted the variational approach described in  \cite{calandra_adiabatic_2010} together with a Wannier interpolating scheme  \cite{marini_epiq_2023} to compute the necessary dynamical linear response functions:  $\sigma_{\text{inter}}$ and $\bar{Z}^{\text{Dyn}}$. Following the procedure described in Ref. \cite{binci_first-principles_2021}, 
  we firstly perform a TD-DFPT calculation at zero frequency and a finite constant $i\eta_\text{reg}$ to represent the $i0^+$ regularization, with  $\eta_\text{reg}=340$ meV.  
  Then, we Wannier interpolate  the previous TD-DFPT calculation for a range of frequencies on a randomized $\bf k$-mesh of $48^3$ points with a Gaussian smearing of 408 meV and $\eta_\text{reg}=340$ meV. We notice that the interband  electronic response  $\sigma_{\text{Inter}}$  does not affect the vibrational resonances in the reflectivity due to the absence of low-energy interband transitions in the frequency range of interest for H$_3$S.
  
  For phonon displacements and frequencies we used the dynamical matrix computed within the Stochastic Self-Consistent Harmonic (SSCHA) approximation in  \cite{bianco_high-pressure_2018}. The refractive index ($n_0$ in equation (\ref{eq_main:R})) has been set to 2.417, accounting for the diamond environment as described in the experimental section of Ref. \cite{capitani_spectroscopic_2017}. 
  
  \subsubsection{Al}
For aluminum we use a plane wave energy cutoff of 25 Ry, a density cut-off of 100 Ry, a Gaussian smearing temperature of 340 meV  and a discrete Brillouin zone sampling using a Monkhorst-Pack mesh of $48^3$ points displaced from the zone center.
The unit cell has FCC symmetry with a lattice parameter of 7.67 (6.53) Bohr  at 0 (100) GPa. For the static and dynamical BECs we follow the same procedure adopted for the H$_3$S case, but with $\eta_\text{reg}=0.14$ meV.
 
  \subsection{\label{si:extdrude}   Electronic response and Extended-Drude calculation} 
  
  To compute the electronic interband response $\sigma_{\text{inter}}(\omega)$, we firstly perform a TD-DFT run following the method illustrated in  \cite{binci_first-principles_2021}.
  Then we employ a Wannier interpolation scheme \cite{marini_epiq_2023} to refine the Brillouin zone sampling and to evaluate the response over a wider range of frequencies. For the intraband contribution  $\sigma_{\text{Drude}}(\omega,\Sigma^{\text{el}})$,  we assess the effect of the scattering via the extended-Drude procedure .
  
  While the theoretical derivation of the extended-Drude calculation is reported in the Method Section, here we describe the actual calculation recipe we implemented following the method described in Ref.   \cite{bickers_infrared_1990}.
  The first step consists in the solution of the Migdal-Eliashberg  \cite{migdal_interaction_1958,eliashberg_interactions_1960} (ME) 
  equations that take into account the effect of electron-phonon and impurity scattering.
  The ME equations are usually defined over the imaginary Matsubara frequencies and under the assumptions of isotropy and localization around the Fermi energy. 
  Within these assumptions, the mass renormalization function $Z_n=Z(i\omega_n)$  and the order parameter $\phi_n =\phi(i\omega_n)$ (identical to zero in the normal phase)  describe the single-particle electronic self-energy written in the Nambu notation as:
  \begin{align}
   \Sigma^{\text{el}}(i \omega_n) = i\hbar \omega_n\left(1-Z_n\right)\tau_0 + \phi_n \tau_1 ,
  \end{align}  
  where $\tau_i$ are Pauli matrices.
  The two functions $Z_n$ and $\phi_n$ can be found solving self-consistently the isotropic ME coupled equations:
    \begin{align}
      \omega_n\left(1-Z_n\right)= & - \frac{\eta_{\text{imp}}}{2} \frac{\omega_n Z_n}{\left[\left(Z_n \omega_n\right)^2+\phi_n^2\right]^{1 / 2}}  -\pi T \sum_m \lambda_{n-m} \frac{\omega_m Z_m}{\left[\left(Z_m \omega_m\right)^2+\phi_m^2\right]^{1 / 2}}
    \\
\phi_n= & \frac{\eta_{\text{imp}}}{2} \frac{\phi_n}{\left[\left(Z_n \omega_n\right)^2+\phi_n^2\right]^{1 / 2}}   +\pi T \sum_m (\lambda_{n-m}-\mu^*) \frac{\phi_m}{\left[\left(Z_m \omega_m\right)^2+\phi_m^2\right]^{1 / 2}}
\end{align}
  where  $i \hbar \omega_n=i(2n+1)\pi k_B T$ labels Fermionic frequencies, $T$ is  the temperature, $\mu^*$ is the Morel-Anderson pseudo-potential \cite{morel_calculation_1962}, $\eta_{\text{imp}}$ is the impurity scattering and
  \begin{equation}
  \lambda_{n-m}=\int_0^{\infty} d \omega \alpha^2 F(\omega) \frac{2 \omega}{\omega^2+\left(\omega_n-\omega_m\right)^2},
  \end{equation}
  with  $\alpha^2 F(\omega)$ the Eliashberg electron-phonon spectral function. 
  We use the \textit{ab initio} $\alpha^2F(\omega)$ comprehensive of quantum and anharmonic effects computed for H$_3$S with the Stochastic Self-Consistent Harmonic (SSCHA) approximation from Ref. \cite{errea_quantum_2016}. 
  Two phenomenological parameters enter in this procedure:
  the impurity scattering rate $\eta_{\text{imp}}$, that we set equal to 135  meV as extrapolated from experimental data \cite{capitani_spectroscopic_2017}, 
  and the Morel-Anderson pseudo-potential, that enters only in the SC phase and that we set to 0.16 in order to reproduce the experimental critical temperature as in Ref.  \cite{capitani_spectroscopic_2017}.
  We then compute the electron-hole propagator over the Matsubara axis
  \begin{align}
  \Pi(i v_m)&=\frac{ (\omega^2_{\text{p}})^{\sigma}}{4\pi }\pi k_B T\sum^{N}_{n=-N} S_{nm}\\
    S_{n m}&=
    \begin{cases}
      \frac{\phi_n^2}{R_n^3}  , & (m = 0)\\
      \frac{1}{R_n} , & (m = -2 n-1))\\
      \frac{\widetilde{\omega}_n\left(\widetilde{\omega}_n+\widetilde{\omega}_{n+m}\right)+\phi_n\left(\phi_n-\phi_{n+m}\right)}{R_n P_{n m}}-
      \frac{\bar{\omega}_{n+m}\left(\bar{\omega}_{n+m}+\bar{\omega}_n\right)+\phi_{n+m}\left(\phi_{n+m}-\phi_n\right)}{R_{n+m} P_{n m}}, & (m \neq 0,-2 n-1)\\
    \end{cases}
  \end{align}  
 with
\begin{align}
& R_n=\left(\widetilde{\omega}_n^2+\phi_n^2\right)^{1 / 2} \\
& P_{n m}=\widetilde{\omega}_n^2-\widetilde{\omega}_{n+m}^2+\phi_n^2-\phi_{n+m}^2
\end{align} 
 being $\tilde{\omega}_n=Z_n \omega_n $ and $i\hbar v_m=i2m \pi k_B T$ a Bosonic Matsubara frequency.
  
  Within this approach the conductivity is obtained as
  \begin{align}
      % \Pi(i v_m)=\pi k_B T\sum^{N_{max}}_n S_{nm} \\
      % I(i v_m)=1 - \Pi(i v_m) \\
      \sigma(i v_m)=i\frac{1}{iv_m}\Pi(i v_m) 
  \end{align}

  We perform an analytical continuation to real frequencies of $\sigma_{reg}(i v_m) = \sigma(i v_m)-\sigma(0)$  using the Pad\'e approximants algorithm described in Ref. \cite{vidberg_solving_1977}.
  The final expression for the intraband contribution to the electronic conductivity is:
  \begin{align}
  \sigma(\omega) = \sigma_{reg}(\omega)+\left(\pi\delta(\omega) +\frac{i}{\omega}\right) \Pi(0) \label{sigmadurudeelia}
  \end{align}
  From the conductivity computed in this way, we extract the factor $I(\omega,\Sigma^{\text{el}})$ defined in equation (\ref{eq:dress_factor_def}) via equation (\ref{eq:sigmadrude}). 
For $\omega\rightarrow 0$, $\sigma_{reg}(\omega)$ remains finite and, in the superconducting state, the optical conductivity, equation (\ref{sigmadurudeelia}), is dominated by the diverging collisionless response of the superconducting condensate fraction given by:
  \begin{align}
  x_{SC} = \frac{4\pi }{ (\omega^2_{\text{p}})^{\sigma}} \Pi(0)
  \end{align}

\begin{figure}[h]%
    \centering
    \includegraphics[width=160mm]{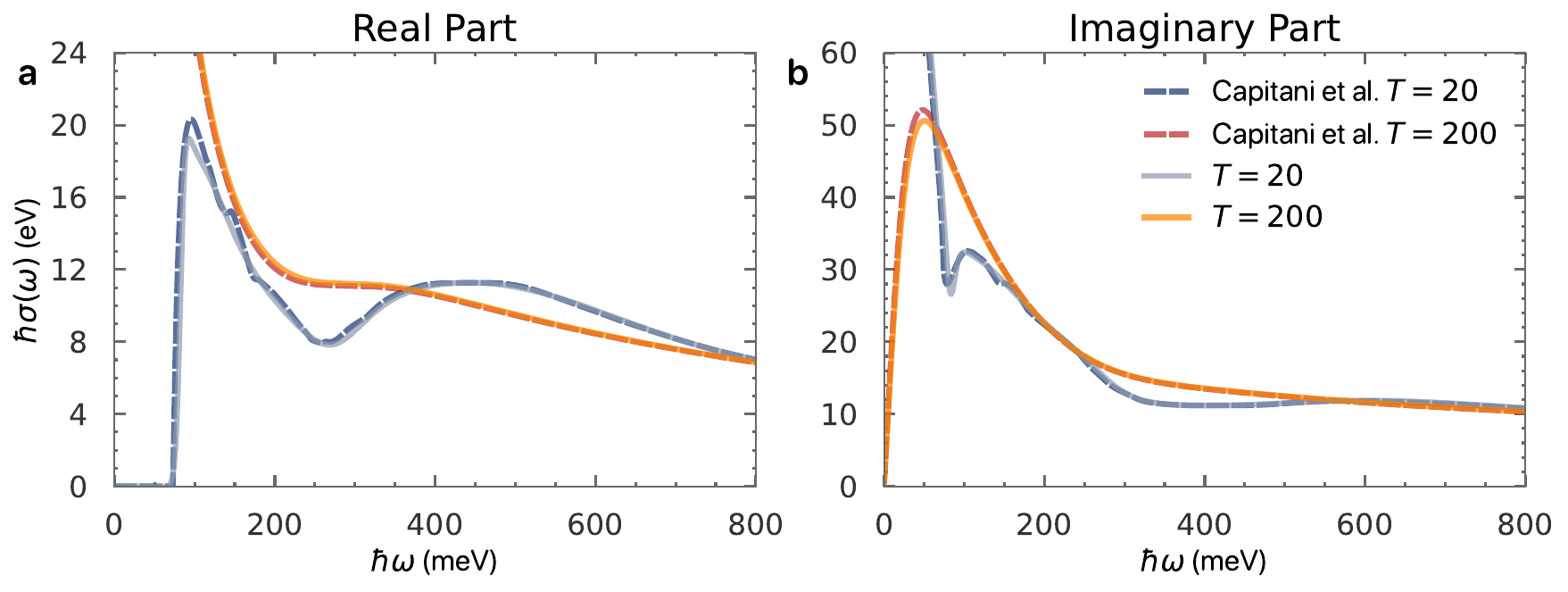}
    \caption{
      \textbf{Validation of the optical conductivity calculation} We validate the approach of Supplementary Section \ref{si:extdrude} by comparing the (\textbf{a}) real and (\textbf{b}) imaginary parts of the optical conductivity with the results obtained in Ref. \cite{capitani_spectroscopic_2017} by employing the direct real-$\omega$-space formalism of Ref. \cite{PhysRevB.43.12804}, a $\alpha^2F(\omega)$ computed at a pressure of 200 GPa and an impurity scattering rate of $\eta_{\text{imp}}=120$ meV.  In both the normal and superconducting phases, the agreement between the calculations is excellent.
    }
    \label{fig:comp1}
  \end{figure}

\begin{figure}[h]%
    \centering
    \includegraphics[width=160mm]{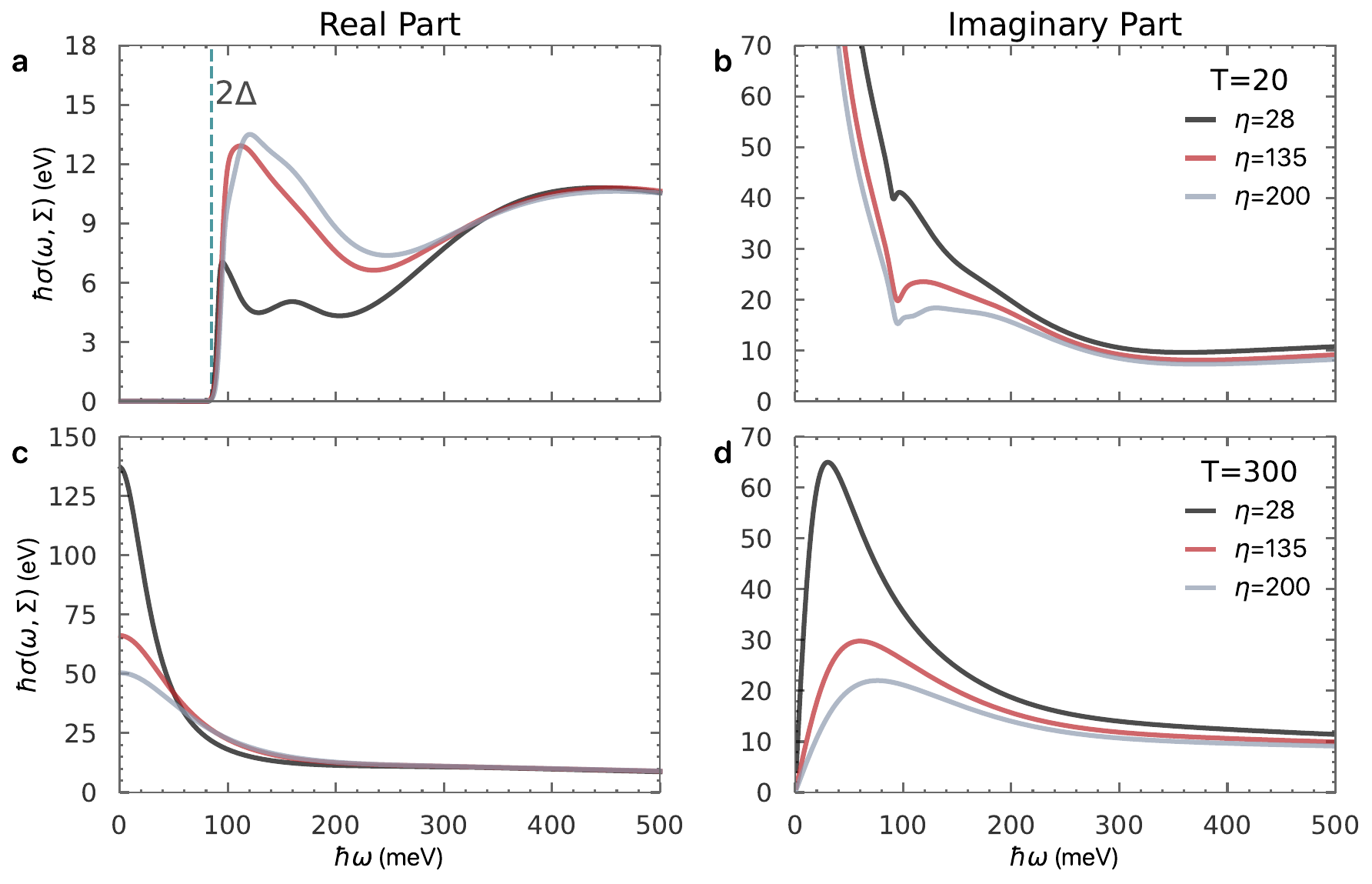}
    \caption{(\textbf{a,c}) Real and (\textbf{b,d}) imaginary parts of the optical conductivity computed as explained in the Supplementary Section \ref{si:extdrude}, for the superconducting phase (\textbf{a,b}) at 20 K and the normal phase at 300 K (\textbf{c,d}), for different impurity scattering rates. 
    }
    \label{fig:comp2}
  \end{figure}

For the ME calculations we set the number of Matsubara frequencies to $N=512$, the temperature to 300 K for the normal (as in the experimental condition) and to 20 K for the SC phase. We use for the numerical analytical continuation 50 Pad\'e coefficients for the conductivity. With this choice of parameters, we compute a conductivity matching the published results  \cite{capitani_spectroscopic_2017}, as shown in Fig. \ref{fig:comp1}. We also investigate the effects of changing the impurity scattering rate in Fig. \ref{fig:comp2}.

We notice here that the presence of anomalous (Gorkov) Green's functions in the SC phase can affect the factorization of vertices and bubble terms of the induced density in equation (\ref{eq:rho_factorized}) and its expansion in the long-wavelength limit leading to equation (\ref{eq:deltaZ_as_prod}) for $\Delta\bar{Z}_{s,\alpha\beta}$. Using Nambu formalism, the electron-hole propagator with bare vertices can be obtained by taking the trace $\mathrm{Tr}\left(\tau_i \hat{G}(k)\tau_i \hat{G}(k+q)\right)$ over a product of generalized Green's functions $\hat{G}(k)$ --- 2$\times$2 matrices in Nambu notation --- and Pauli matrices $\tau_i$. The (bare) vertex structure is irrelevant in the normal phase, where anomalous Green's functions vanish and $\hat{G}$ are diagonal matrices: in this case, the relation between BECs and the isotropic Extended-Drude model is exact within the approximations discussed in Methods. When neglecting vertex corrections, the electron-hole propagator entering in the optical conductivity is obtained with $\tau_0$ in the bare vertices\cite{bickers_infrared_1990}, while the induced charge is given by a polarization bubble with $\tau_3$ structure\cite{zeyher_zphysB1990}. The factorization and long-wavelength expansion of such vertices in the interacting SC state cannot be performed in the Kohn-Sham DFPT context adopted for the normal phase and described in Methods. Rather, a full quantum-field many-body approach would be required to first evaluate the fully interacting electron-hole propagator at any momentum $\bm q$. In order to satisfy the Ward identity --- and hence the continuity equation --- vertex corrections at finite momenta as well as contributions from SC collective modes (SC amplitude and, more importantly, SC phase modes)\cite{schrieffer_book, benfatto_prb2016, hirschfeld_prb2017} need to be included before carrying on the expansion of the induced density in the long-wavelength limit and identifying the intraband contribution to BECs analogous to equation (\ref{eq:deltaZ_as_prod}). Instead of  undertaking this quite formidable task, we opted for a conservative choice in the SC phase: we assumed that the adimensional factor ${I}(\omega,\Sigma^{\text{el}})$ entering in the dressing of $\bar{Z}_{s,\alpha\beta}$ can still be expressed as in equation (\ref{eq:BECwithDrude1}), i.e., as a function of the damped and undamped Drude-like contribution to the optical conductivity evaluated in this section, for which the neglect of vertex corrections is generally assumed as a reasonable approximation \cite{bickers_infrared_1990}. We also investigated the electron-hole propagator with $\tau_3$ bare vertices, which leads to much more enhanced Born effective charges at the SC gap resonance and thus to a much stronger spectroscopic signature of the superconducting transition compared to that displayed in \ref{fig:R_excitations} of the paper.
  
  \subsection{\label{si:vibr} Lattice vibrations}
  Dynamical matrices are response functions describing the phonon modes, 
  i.e., the collective ionic motions of the crystal, characterized by a frequency $\omega_{\mu}$, a displacement $\mathbf e_{\mu}$ and a lifetime $\gamma_{\mu}$ for each mode $\mu$.
  Thanks to the method of Ref.  \cite{calandra_adiabatic_2010} 
  and the extended-Drude for phonons described in 
  Refs.  \cite{maksimov_nonadiabatic_1996,saitta_giant_2008} we find that the phonons in H$_3$S are weakly affected by scattering and dynamical effects. As a matter of facts, scattering affects dynamical matrices in perfect analogy with BECs. It follows that scattering effects are negligible when the difference between the static and dynamic limit (adiabatic and non-adiabatic in the notation of ref. \cite{saitta_giant_2008} ) of the phonon frequencies is small. 
  Hence, we use the adiabatic dynamical matrices retrieved via the SSCHA  that accounts for quantum and anharmonic effects.
  % We adopt the lifetime due to phonon-phonon scattering obtained from SSCHA.
  In   \ref{tab:phonon} we report the data used in our calculations.
  Note that we set a different lifetime for the low frequency phonon ($\omega_{\mu}=84$ meV), higher than the one obtained from SSCHA. 
  Indeed, the latter is just a lower bound since does not take into account the electron-phonon or the impurity but only the phonon-phonon scattering.
  We choose a reasonable value considered also graphical requirements.

  We quantify the strength of the oscillation by the `vibrational plasma frequency' defined as 
  \begin{align}
    (\Omega^2_{\text{p},\mu})_{\alpha\beta}(\omega_{\mu},\Sigma^{\text{el}})=\frac{e^2}{\Omega}\sum^{\omega_{\mu'}=\omega_{\mu}}_{\mu'} d^{\mu'}_{\alpha}(\omega_{\mu'},\Sigma^{\text{el}})d^{\mu'}_{\beta}(\omega_{\mu'},\Sigma^{\text{el}}),
  \end{align}
  where we sum over degenerate modes, and presented for H$_3$S in \ref{tab:phonon} with its unique independent component $ (\Omega^2_{\text{p},\mu})_{\alpha\beta}=\delta_{\alpha\beta}\Omega^2_{\text{p},\mu}$. We report in the table all the ingredients required for computing the reflectivity at the phonon frequencies and entering in equations (\ref{eq_main:R}), (\ref{eq:epsilontot}) and (\ref{eq:chivibr}).
  
  \begin{table}[h]
    \begin{tabular}{cc|cc|cc|c|c|c|c}
    \multicolumn{6}{c|}{ } & \multicolumn{2}{c|}{T=20 K}& \multicolumn{2}{c}{T=300 K}\\
      	 $\hbar \omega_{\mu}$   	&	 $\hbar \gamma_{\mu}$  &	Irreps 	&	  Activity &
      $\hbar\Omega^{\text{Stat}}_{\text{p},\mu}  $ &	 $\hbar\Omega^{\text{Dyn}}_{\text{p},\mu}(\omega_{\mu})  $ &	 $\hbar\Omega^{\text{}}_{\text{p},\mu}(\omega_{\mu},\Sigma^{\text{el}})  $ 	&	$1+4\pi\chi^{\text{el}}(\omega_{\mu},\Sigma^{\text{el}})$  &  $\hbar\Omega^{\text{}}_{\text{p},\mu}(\omega_{\mu},\Sigma^{\text{el}})  $ 	&	$1+4\pi\chi^{\text{el}}(\omega_{\mu},\Sigma^{\text{el}})$  \\
      \hline
      	 84 	& $1$   &		$ T_{1u}  $	&	 InfraRed &	 $280$  & $751+i6$   &	$96-i5$   & $-4375+i31$    & $106+i159$ & $-4141+i3967$  \\
      	 148 	& $2.3$ &		$ T_{1u}  $	&	 InfraRed &	 $80$   & $961+i21$    &	$286-i79$ & $-1826+i928$   & $271-i120$ & $-1683+i1307$  \\
      	 181 	& $2.2$ &	   $ T_{2u}  $	&	          &          &              &	          & $-1294+i610$   &            & $-1153+i923$  \\
    \end{tabular} 
    \caption{\textbf{Table of vibrational response for H$_3$S at 150 GPa.} 
    We report the phonon descriptors retrieved via the SSCHA calculation  \cite{bianco_high-pressure_2018} (frequency, linewidth, irreducible representation and spectroscopic activity). The corresponding vibrational plasma frequencies obtained with the BECs in the dynamic, static and damped regime are presented at 20 K and 300 K.
    Similarly, the last column lists the value of the electronic contribution to the dielectric tensor at the phonon frequency. 
    The energy unit adopted in the table is meV. Notice that the vibrational plasma frequency and the dielectric response for the lowest phonon at 20 K are almost real.
    }
    \label{tab:phonon}
  \end{table}

  \subsection{\label{si:becs} Numerical calculation of Born Effective Charges}
  There are two equivalent ways of calculating BECs: either as the finite-difference derivative of the induced charge density equation (\ref{eq:defzbar}), (\ref{eq:bubblerho}) or from the direct evaluation of the linear response expressions  (\ref{eq:dyn_bec}), (\ref{eq:deltaZ_as_prod}).
    After checking that both methods coincide, we choose different approaches for the dynamic and static limits.
 
 We remark here that both procedures require a preliminary DFPT computation at $\mathbf{q}=\mathbf{\Gamma}$
 in order to determine the Fermi shifts necessary to the definition of the unscreened deformation potential (\ref{eq:def_pot}), unless they are zero by symmetry.  
  
  \subsubsection{Dynamic Born Effective Charges} 
    We implement the method described in  \cite{binci_first-principles_2021} for computing dynamical BECs with a direct  TD-DFPT calculation that is equivalent to equation (\ref{eq:dyn_bec}).
  In the actual calculation, the discrete sampling of the Brillouin zone forces to use a finite value for the regularization parameter $i0^+$. We employ the smallest value adequate to remove non-physical behaviors due to sampling finiteness.
    Since the explicit form of dynamical BECs does not depend on any derivatives with respect to the momentum $q$, it is computationally more efficient to 
    perform a single TD-DFT run at $\mathbf{q}=\mathbf{\Gamma}$ rather than studying the long-wavelength limit of the induced density stripped of its intraband contribution.
    However, as a consistency check, we verify  the agreement of these two computational schemes.
    Subsequently, we extend the frequencies range and refine the Brillouin sampling with the same Wannier interpolation scheme  \cite{marini_epiq_2023} used for the optical conductivity.
    
  \subsubsection{Static Born Effective Charges}
\begin{figure}[h]
    \centering 
    \includegraphics[width=.3\textwidth]{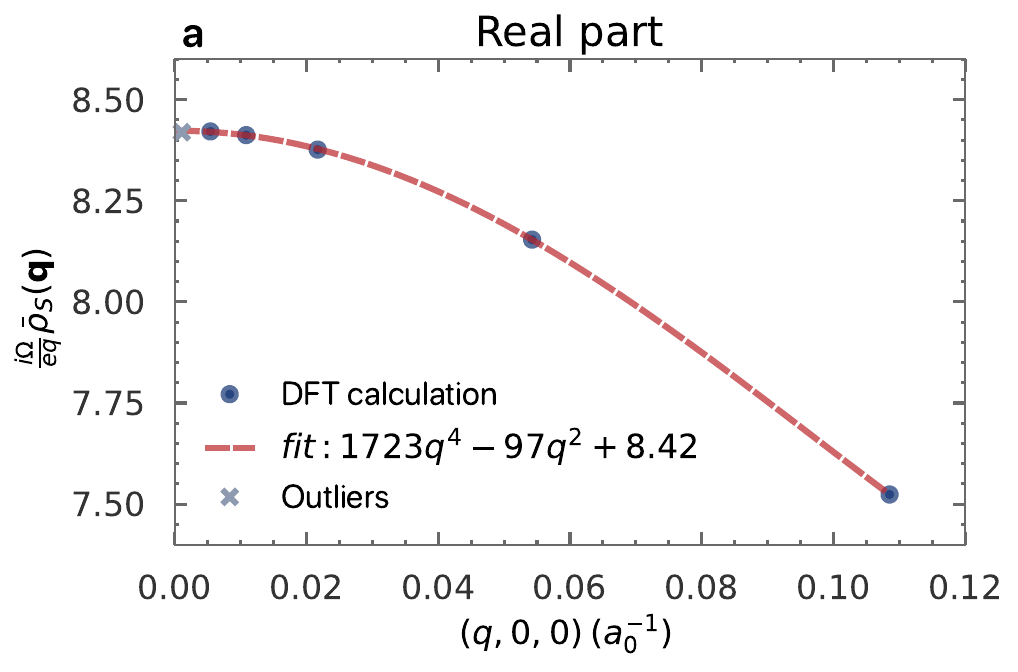}~
    \includegraphics[width=.3\textwidth]{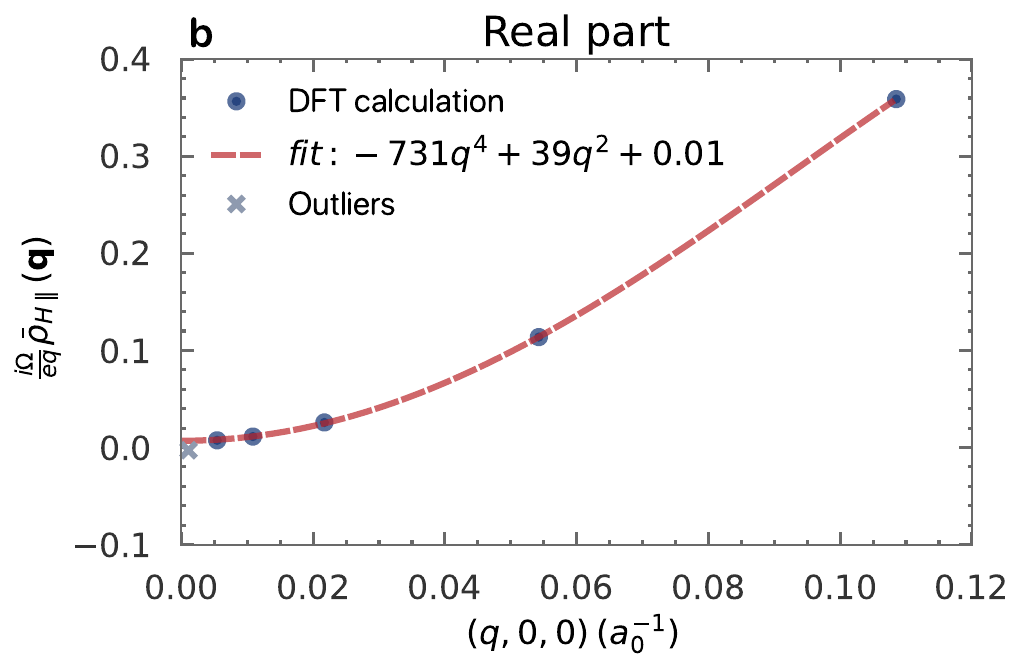}~
    \includegraphics[width=.3\textwidth]{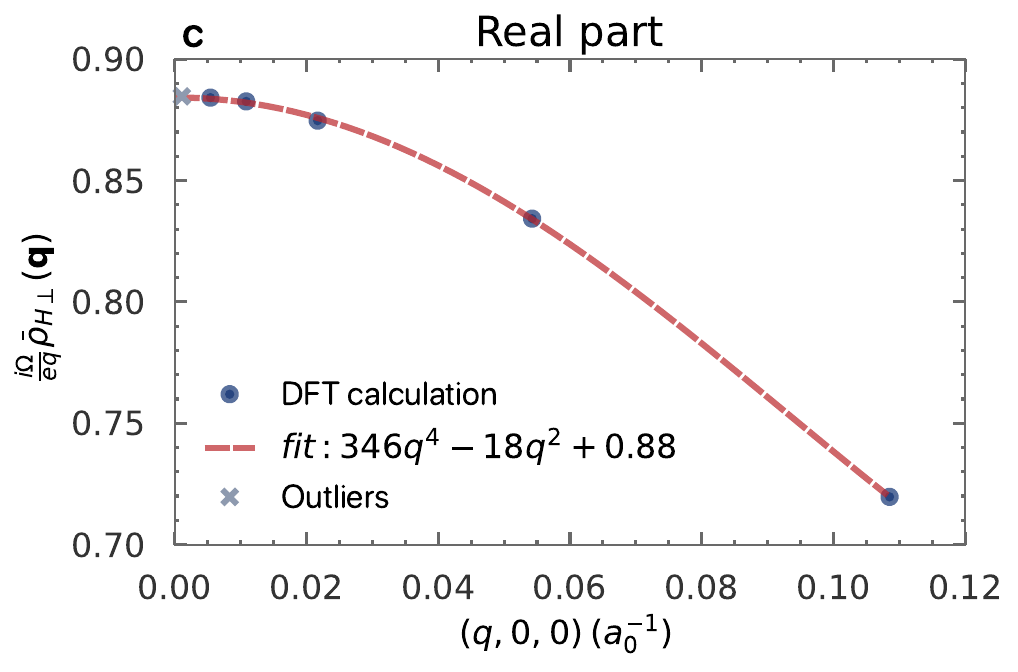}
    \caption{\textbf{The long-wavelength limit of the induced charge density.}  
    With DFPT calculations (blue dots) we study the ratio of the induced density (equation (\ref{eq:bubblerho})) to the wavelength module for each independent component of H$_3$S at 150 GPa: $S$, $H\parallel$ and $H\perp$, dispayed in (\textbf{a}), (\textbf{b}) and (\textbf{c}), respectively.  The dashed red line depicts the polynomial fit of DFPT calculations, whose coefficients are presented in the key; the extrapolated value of the fit at $q=0$ corresponds to the static BECs. The gray crosses indicates DFPT calculation not reliable  for the presence of numerical noise at small momentum and thus not considered in the polynomial fit.
    The imaginary part is negligible ($\sim 10^{-7}$). 
    }
    \label{fig:drho_limit}
  \end{figure}

  The evaluation of static BECs requires the computation of a derivative with respect to the exchanged momentum $\mathbf{q}$: whether for evaluating the second term in equation (\ref{eq_main:singul_step}) to be inserted in equation (\ref{eq:bec_at_1st}), namely
\begin{align}
     \frac{2}{N_\mathbf{k}}\sum^{N_\mathbf{k}}_{\mathbf{k},l} f^{'}(\varepsilon_{\mathbf{k}}^{l})
    \bigg\langle{u^{l}_{\mathbf k}}\bigg|
    { \frac{\partial \bar{V}^{\mathbf{\Gamma} }_{s\beta }}{i\partial q_{\alpha}}   }\bigg|
    {u^{l}_{\mathbf k} } \bigg\rangle, 
    \label{eq:deriv_term}
\end{align}
    or in order to calculate the complete BECs as the long-wavelength limit of the induced density $\bar{\rho}_{s\beta}(\mathbf q)$, equations (\ref{eq:defzbar}), (\ref{eq:bubblerho}).
   % \fmau{collegati a equazioni del main, invece della 4-5, usi 14-15 per comodita' visto che comunque devi calcolare la derivata di V(u) in dq del secondo termine della eq. 20}
   We chose the latter scheme and implemented in the Quantum Espresso package  \cite{giannozzi_quantum_2009,giannozzi_advanced_2017} the necessary modifications for the unscreening procedure described in the method section.
    As shown in \ref{fig:drho_limit}, we  obtain static BECs as the limiting value of the ratio between the induced density and the exchanged momentum estimated by the interpolation of  DFPT calculations at small momentum $\mathbf{q}$.
    Taking advantage of TRS, we employ a fourth order polynomial as fitting function.
    At very small momenta the results are affected by numerical noise and they should be discarded.
    On the other hand, the smallest but reliable point gives already the asymptotic $|{\bf q}|=0$ result and could be used alone for the estimation of the limiting value.
    
\begin{figure}[h]
    \centering
    \includegraphics[width=.3\textwidth]{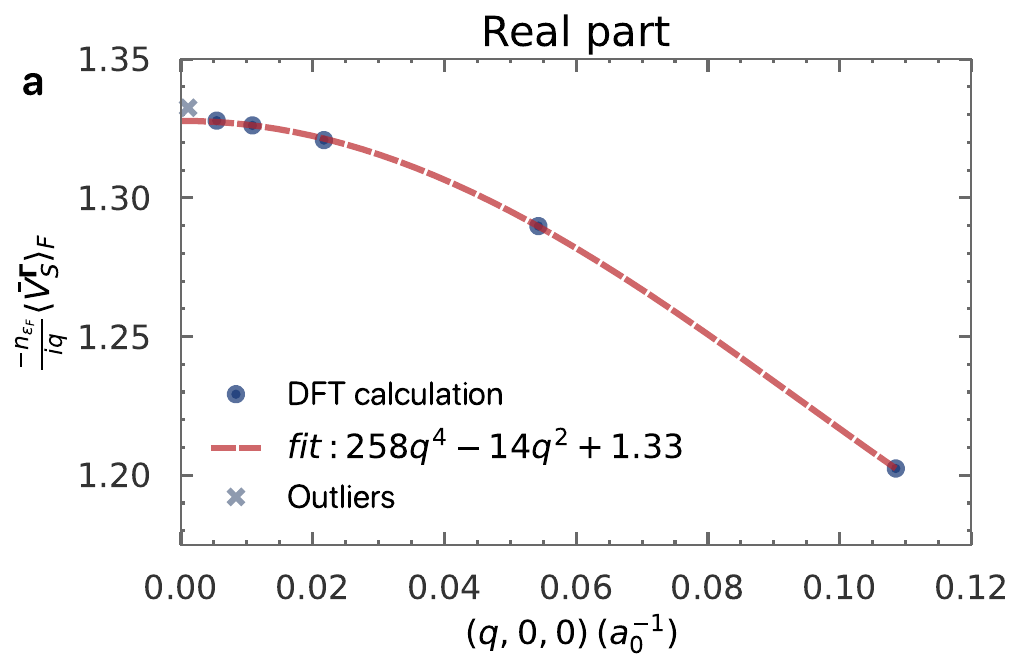}~
    \includegraphics[width=.3\textwidth]{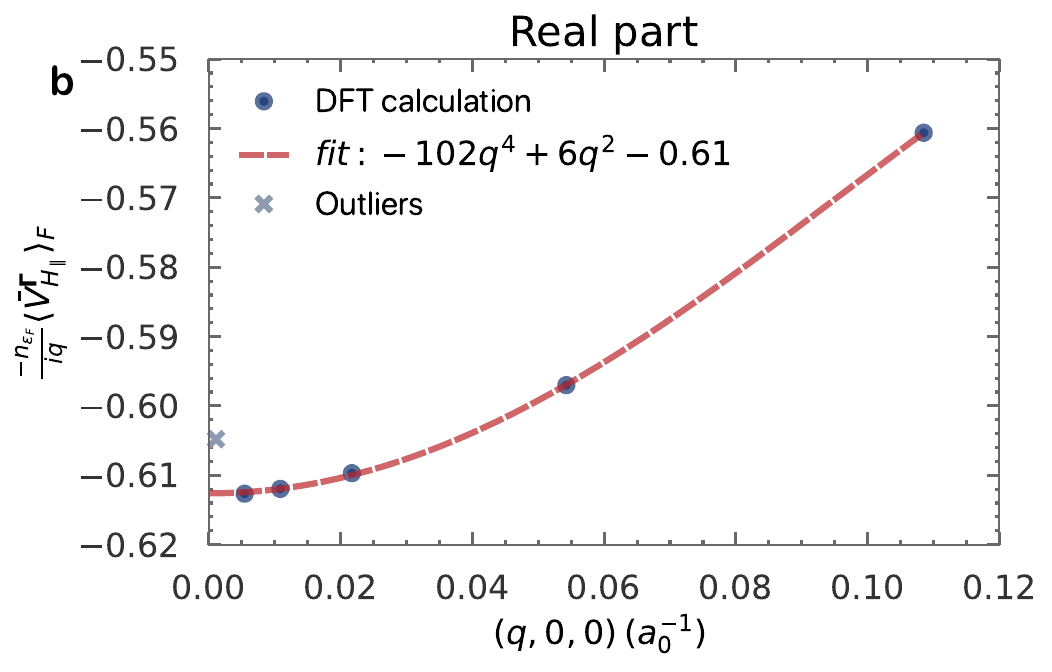}~
    \includegraphics[width=.3\textwidth]{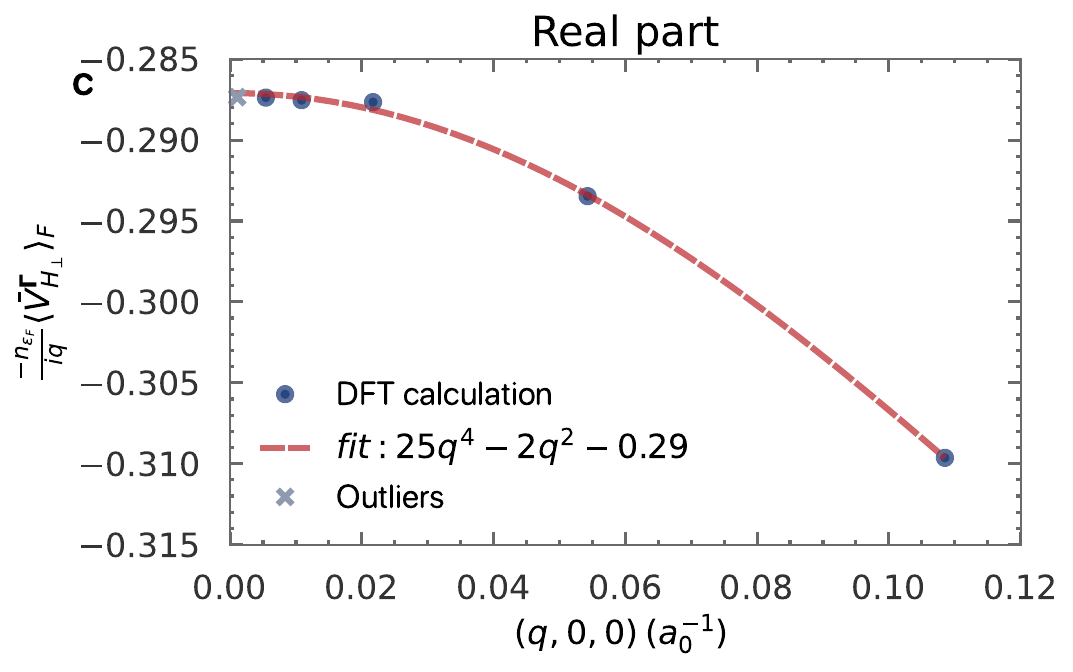}
    \caption{\textbf{The long-wavelength limit of the deformation potential at the Fermi level.}   
    We obtain the derivative of the deformation potential averaged over the Fermi surface with respect to the momentum from finite differences, interpolating different DFPT calculations (blue dots) with a polynomial fit (red dashed curve).
    Each panel shows one independent component $S$, $H\parallel$ and $H\perp$ [(\textbf{a}),(\textbf{b}) and (\textbf{c}), respectively].
    The gray crosses indicates DFPT calculation not reliable  for the presence of numerical noise at small momentum and thus not considered in the polynomial fit.
    The imaginary part is negligible ($\sim 10^{-8}$). 
    }
    \label{fig:g_F}
  \end{figure}

    We adopt an analogous procedure to compute the contribution (\ref{eq:deriv_term}) required for the validation of the static sumrule  \ref{fig:g_F}.
    
  \begin{table}[h]
  \centering
    \begin{tabular}{cc|c|c|c|c}   
  && $Z^S $ & $ Z_{\parallel}^H $ & $ Z_{\perp}^H $ & $\sum_s $ \\
      \midrule
  \midrule 
130 GPa &  $\bar{Z}^{\text{Dyn}} (i0^+)$ & $-1.483$& $3.866$& $-0.184$& $2.015$\\
        &  $\bar{Z}^{\text{Stat}}$ & $8.060$& $0.111$& $0.848$& $9.867$\\
        &  $- n_{\varepsilon_f}\langle\frac{\partial \bar{V}^{\mathbf{\Gamma} }  }{i\partial q }\rangle_F$& $1.215$& $-0.600$& $-0.304$& $0.007$\\[1ex]
\midrule 
150 GPa & $\bar{Z}^{\text{Dyn}} (i0^+)$ & $-1.306$& $3.744$& $-0.205$& $2.028$\\
        & $\bar{Z}^{\text{Stat}}$ & $8.423$& $0.007$& $0.884$& $10.199$\\
        & $- n_{\varepsilon_f}\langle\frac{\partial \bar{V}^{\mathbf{\Gamma} }  }{i\partial q }\rangle_F$& $1.328$& $-0.613$& $-0.287$& $0.141$\\[1ex]
\midrule 
200 GPa & $\bar{Z}^{\text{Dyn}} (i0^+)$ & $-0.758$& $3.329$& $-0.261$& $2.048$\\
        & $\bar{Z}^{\text{Stat}}$ & $9.156$& $-0.170$& $0.970$& $10.926$\\
        & $- n_{\varepsilon_f}\langle\frac{\partial \bar{V}^{\mathbf{\Gamma} }  }{i\partial q }\rangle_F$& $1.584$& $-0.640$& $-0.253$& $0.437$\\[1ex]
\end{tabular}
    \caption{\textbf{Individual contribution to BECs sumrule for H$_3$S.}
    For the three pressure values (130, 150 and 200 GPa) at 300 K, we report the contribution to the BECs sumrules due to each independent component $Z^S $, $ Z_{\parallel}^H $, $ Z_{\perp}^H $. In particular, for each pressure, in the first two rows we list BECs in the dynamic and static limits. In the third row instead we analyze the term $- n_{\varepsilon_f}\sum_s\langle\frac{\partial \bar{V}^{\mathbf{\Gamma} }  }{i\partial q }\rangle_F$ appearing in the r.h.s of equation (\ref{eq_main:adiab_sumrule}), whose small summed value comes from a cancellation over different components. The computational parameters adopted are given in Section \ref{si:calc_param}.
    Note that in the case of elemental Al (a cubic system with one atom per unit-cell) the BECs at different pressures coincide with the sum rule values given in \ref{tab:h3s_al_sumrule}.
    }
    \label{tab:h3s_sumrule_comps}
  \end{table}

  \clearpage
  \bibliography{h3s_prl}% Produces the bibliography via BibTeX.

\end{document}